\documentclass[12pt]{iopart}
\usepackage{iopams} 
 
\usepackage{color}
\usepackage{ulem}
\usepackage{hyperref}
\usepackage{graphicx}
\usepackage{subfig}

\begin{document}

\title{Nonlinear edge localized mode with impurity seeding in CFETR hybrid scenario}

\author{Shiyong Zeng}
\address{State Key Laboratory of Advanced Electromagnetic Technology, International Joint Research Laboratory of Magnetic Confinement Fusion and Plasma Physics, School of Electrical and Electronic Engineering, Huazhong University of Science and Technology, Wuhan, Hubei 430074, China}
\vspace{10pt}

\author{Ping Zhu}
\address{State Key Laboratory of Advanced Electromagnetic Technology, International Joint Research Laboratory of Magnetic Confinement Fusion and Plasma Physics, School of Electrical and Electronic Engineering, Huazhong University of Science and Technology, Wuhan, Hubei 430074, China}
\address{Department of Nuclear Engineering and Engineering Physics, University of Wisconsin-Madison, Madison, Wisconsin 53706, USA}
\ead{zhup@hust.edu.cn}
\vspace{10pt}

\vspace{10pt}
\begin{indented}
\item[]\today
\end{indented}

\begin{abstract}
A critical challenge for operating fusion burning plasma in high-confinement mode lies in mitigating damage caused by edge localized modes (ELMs). While impurity seeding has been experimentally validated as a reliable and effective ELM mitigation technique, its underlying physics remains insufficiently understood and requires further clarification. Through nonlinear magnetohydrodynamic (MHD) simulations, this work reproduces key features of natural ELM crash and reveals its trigger mechanism. Impurity seeding significantly affects nonlinear ELM dynamics by inducing local and global modifications to the pedestal pressure profile, driving high-$n$ ballooning mode instabilities that govern ELM crash. Two critical control parameters---impurity density level and poloidal seeding location---are systematically investigated, which play key roles in the ELM crash onset timing and the resulting energy loss magnitude.
\end{abstract}

\title[Nonlinear ELM with impurity seeding in CFETR plasma]{}

\maketitle

%
%
%
%
%

\section{Introduction}
\label{Sec:introduction}

The operation scenario for next-generation tokamaks aiming to achieve fusion burning plasmas fundamentally relies on the high-confinement mode (H-mode) \cite{Wagner_Hmode_PhysRevLett.49.1408}, which is characterized by the formation of a narrow transport barrier at the plasma edge. The consequent periodic burst instability at the pedestal region, known as the edge localized mode (ELM), exhibits a dual nature: while preventing impurity accumulation in the plasma core, it concurrently releases intense heat and particle fluxes within milliseconds, posing severe threats to the lifetime of plasma-facing components (PFCs).
Experimental data reveal that the crash size of the most hazardous type-\uppercase\expandafter{\romannumeral1} ELM shows strong correlation with pedestal plasma collisionality $\nu^{*}_{ped}$ \cite{LOARTE2003962}, quantified by $\Delta W_{ELM}/W_{th}$, where $\Delta W_{ELM}$ represents the energy loss per ELM crash and $W_{th}$ denotes the plasma stored thermal energy. Projections indicate ELM sizes of approximately $19\%$ ($\Delta W_{ELM}\approx22$ MJ) in ITER \cite{LOARTE2003962} and $17\%$ ($\Delta W_{ELM}\approx51$ MJ) in China Fusion Engineering Test Reactor (CFETR) plasmas ($\nu^{*}_{ped}\sim0.05$\footnote{$\nu^{*}_{ped}=\epsilon^{-3/2} Rq_{95}/\lambda_{e,e}$, where $\epsilon$, $R$, and $q_{95}$ are the inverse aspect ratio, major radius, and safety factor at the $95\%$ flux surface, respectively. The electron-electron Coulomb collision mean free path length $\lambda_{e,e}=1.44\times10^{23}T_e^2/\left(n\ln\Lambda\right)$ with $\ln\Lambda=14.9-0.5\ln\left(n/10^{20}\right) +\ln T_e$ (units: m, keV, m$^{-3}$) follows the formulation in the appendix of Wesson J. 2011 Tokamaks 4th edn (Oxford: Oxford University Press).}) \cite{Zhuang_2019}. However, engineering constraints mandate $\Delta W_{ELM}<1$ MJ m$^{-2}$ for PFC protection \cite{Ueda_2017}.
Substantial progress has been achieved in elucidating ELM physics through systematic investigations \cite{Wade_PhysRevLett.94.225001,Kirk_PhysRevLett.96.185001,Lang_2013_overview,Loarte_2014,Leonard_2014,Kirk_2014,Huijsmans_2015,Xiao_2020_HL_2A,Xu_2023_EAST}. The prevailing paradigm attributes type-\uppercase\expandafter{\romannumeral1} ELM to ideal magnetohydrodynamic (MHD) peeling-ballooning mode (PBM) instabilities \cite{Snyder_2002}. Recent advances reveal that resistive PBMs develop prior to reaching the ideal MHD stability limit when finite plasma resistivity is considered \cite{Kleiner_2021,Nyström_2022}, yielding better agreement with experimental observations. Complementary mechanisms---including the nonlinear explosive growth of ballooning modes \cite{Wilson_PhysRevLett.92.175006,Ham_PhysRevLett.116.235001} and three-wave coupling phenomena observed in D\uppercase\expandafter{\romannumeral3}-D plasmas \cite{Diallo_PhysRevLett.121.235001}---offer alternative frameworks for interpreting ELM dynamics.





Building upon the fundamental understanding of natural ELM, extensive research efforts have focused on mitigating type-\uppercase\expandafter{\romannumeral1} ELM by converting large-amplitude/low-frequency crash into small-amplitude/high-frequency event or ELM-free regime, as demonstrated through diverse techniques across multiple devices: including impurity seeding using main ion or noble gas on JET \cite{Beurskens_2008}, ASDEX Upgrade \cite{Schneider_2014}, and EAST \cite{Li_2020}; supersonic molecular beam injection (SMBI) on EAST \cite{LIN2022127988}, HL-2A \cite{Xiao_2012,Zhong_2019}, and KSTAR \cite{Xiao_2014}; laser blow-off metal injection on HL-2A \cite{Zhang_2018,Xiao_2021}; lithium granule/boron powder injection on EAST \cite{Sun_2021b,Sun_2021,Ye_2021}; and pellet injection on D\uppercase\expandafter{\romannumeral3}-D \cite{Baylor_PhysRevLett.110.245001,Baylor_2013,Wilcox_2022}, JET \cite{Lang_2011,Lang_2013}, and ASDEX Upgrade \cite{Lang_2004,Kocsis_2007}.
Furthermore, numerical simulations have significantly advanced the understanding of impurity seeding ELM physics \cite{Huijsmans_2015}.
BOUT++ simulations demonstrate quantitative correlations between ELM size and pedestal resistivity \cite{Xu_PhysRevLett.105.175005,Dudson_2011,Wu_2018}.
M3D-C1 simulations reveal a pronounced dependence of the pellet mass threshold on poloidal injection location for ELM triggering \cite{Wingen_2024}.
JOREK simulations reproduce multi-ELM cycles using a reduced MHD model \cite{Orain_PhysRevLett.114.035001,Cathey_2020,Cathey_2022}, incorporating a pellet ablation model that demonstrates localized high-density plasmoid can drive ballooning-type instabilities \cite{Huysmans_2009}. Moreover, comparative analyses between natural and pellet triggered ELMs have been conducted \cite{Futatani_2020,Cathey_2021}, and the experimentally observed lag time in pellet triggered ELM has been reproduced \cite{Futatani_2021}.
NIMROD simulations demonstrate excellent agreement with ELITE and GATO codes in benchmark studies \cite{Burke_2010}, and enable two-fluid analysis of edge-localized instabilities \cite{Brennan_2006,Sovinec_2007}. Subsequent studies further examine edge current density modulation effects on pedestal stability \cite{PZhu_2012} and provide insights into ELM suppression in lithium-coated NSTX experiments \cite{Banerjee_2017}. Additional applications include quantifying enhanced plasma resistivity stabilization effects on peeling-ballooning instabilities in EAST \cite{Lin_2020} and modeling edge harmonic oscillations in D\uppercase\expandafter{\romannumeral3}-D Quiescent H-mode discharges \cite{Pankin_2020}.
Simulation studies have continuously advanced, such as developing models for kinetic effects that describe turbulent and neoclassical transport in the pedestal region, and creating more complete scrap-off layer/divertor models that incorporate the interaction of neutral particles and plasma.

Despite extensive experimental and simulation advancements, the physics underlying impurity seeding mechanisms for ELM triggering requires further investigation \cite{Zhang_2018,Xiao_2021,Sun_2021,Wilcox_2022,Lang_2011,Lang_2013,Lunsford_2018}. Previous studies propose that the local plasmoid induced by pellet ablation cloud can reinforce the pedestal pressure gradient, driving ballooning mode instabilities that trigger ELM \cite{Baylor_PhysRevLett.110.245001,Baylor_2013,Kocsis_2007}. However, JOREK simulations indicate that localized pressure gradient enhancement alone cannot fully account for ELM triggering \cite{Futatani_2014}. Lithium coating experiments on NSTX demonstrate significant impacts of entire pedestal profiles \cite{Maingi_PhysRevLett.107.145004}, with the pedestal density identified as a critical parameter \cite{LIN2022127988,Harrer_2018}. ASDEX Upgrade experiments further show that outward pedestal density shifts destabilize PBMs, while inward shifts stabilize them \cite{Dunne_2017}.
Regarding ELM size-frequency correlations, experiments exhibit reduced ELM size alongside increased frequency, in agreement with the assumption $f_{ELM}\times\Delta W_{ELM}\approx const$, where $f_{ELM}$ denotes ELM frequency. By contrast, lithium granule injection induces reductions in both ELM size and frequency \cite{Sun_2021b}. Neon impurity fractions exhibit non-monotonic effects on ELM frequency \cite{Zhong_2019}, a trend also observed in JET \cite{Beurskens_2008}.

To advance our understanding on the ELM mitigation via impurity seeding, this study investigates the nonlinear dynamics of ELM with impurity seeding using 3D nonlinear MHD simulations. The effects of impurity seeding---specifically the density perturbations from impurity ionization and the temperature perturbations from radiative cooling---are examined, revealing that impurity induced pedestal density perturbations play a critical role in ELM dynamics under the scenario studied. Two key impurity seeding parameters demonstrating significant impacts on the ELM crash onset and size are systematically analyzed.

The paper is organized as follows: Section \ref{Sec:simulation model and equilibrium} introduces the simulation model, the numerical setup, and the equilibrium configuration. Section \ref{Sec: nonlinear dynamic of ELM crash} characterizes the natural ELM crash mechanism and the energy loss in absence of impurity seeding. Section \ref{Sec: Impurity seeding} quantifies the impurity density level and poloidal seeding location effects on ELM dynamics. Section \ref{Sec: discussion} gives discussion, and Section \ref{Sec: conclusion} summarizes key findings.

\section{Simulation model and equilibrium}
\label{Sec:simulation model and equilibrium}

\subsection{Simulation model and setup}
The numerical simulations are conducted using the NIMROD code \cite{Sovinec2004}. Our implementation employs a full MHD model that integrates an atomic physics module adapted from the KPRAD code \cite{KPRAD,Izzo_2013}. The governing equations are formulated as:
\begin{eqnarray}
	\frac{\partial n_{i,Z}}{\partial t} + \nabla \cdot \left(n_{i,Z}\bi{V}\right) = \nabla \cdot (D \nabla n_{i,Z}) + S_{ion/rec},
	\label{eq:contiune2}
	\\
	\rho \left( \frac{\partial\bi{V}}{\partial t} + \bi{V} \cdot \nabla \bi{V} \right) = - \nabla p + \bi{J}\times\bi{B} + \nabla\cdot\left(\rho\nu\nabla\bi{V}\right),
	\label{eq:momentum}
	\\
	\frac{n_e}{\Gamma-1}\left( \frac{\partial T_e}{\partial t} + \bi{V} \cdot \nabla T_e\right) = -n_e T_e \nabla \cdot \bi{V} - \nabla \cdot \bi{q} + Q,
	\label{eq:temperature}
	\\
	\bi{q} = -n_e[\kappa_{\parallel} \bi{b}\bi{b} + \kappa_{\perp} (\mathcal{I} - \bi{b}\bi{b})] \cdot \nabla T_e,
	\label{eq:heat_flux}
	\\
	\frac{\partial \bi{B}}{\partial t} = \nabla\times\left( \bi{V}\times\bi{B}\right) - \nabla\times\left( \eta\bi{J}\right),\quad
	\label{eq:Faraday law}
	\mu_0\bi{J} = \nabla\times\bi{B}.
	\label{eq:Ampere law}
\end{eqnarray}
Here, $n_i$, $n_e$, and $n_Z$ denote the main ion, electron, and impurity ion number densities, respectively; $\rho$, $p$, and $T_e$ represent the plasma mass density, pressure, and electron temperature; $\bi{V}$, $\bi{J}$, and $\bi{q}$ correspond to the velocity, current density, and heat flux; $\bi{B}$ is the magnetic field; $\bi{b}=\bi{B}/B$ defines the unit magnetic field vector; $\mathcal{I}$ represents the unit tensor; $S_{ion/rec}$ denotes the density source; $Q$ corresponds to the energy source; $D$, $\nu$, $\eta$, $\kappa_{\parallel}$, and $\kappa_{\perp}$ characterize the plasma diffusivity, kinematic viscosity, resistivity, parallel and perpendicular thermal conductivity, respectively, and $\Gamma = 5/3$ is the adiabatic index.

The KPRAD module calculates the impurity radiation power with real-time updates of impurity and plasma densities. The energy source term $Q$ comprises contributions from impurity ionization, recombination, line radiation, bremsstrahlung, and Ohmic heating. All particle species adhere to the single-temperature $T_e$ approximation, which assumes instantaneous thermalization across species due to collision timescales being much shorter than the ELM crash timescale $\tau_{ELM}$ in the edge pedestal region under impurity seeding. For instance, the electron collision time $\tau_e=6.4\times10^{14}(T_e^{3/2}/Z^2n_e)\simeq0.002$ ms $\ll\tau_{ELM}\simeq0.1$ ms, which are estimated for ion charge number $Z=5$, $T_e=2.5$ keV, and $n_e=5\times10^{19}$ m$^{-3}$. Ionization and recombination processes are modeled for both plasma ion and multi-charged impurity ions, with their contributions integrated into the density source term $S_{ion/rec}$. Plasma ions consist of deuterium and are initially fully ionized; neutral deuterium generated via recombination after impurity seeding accounts for only a small fraction of the plasma ion bulk, rendering it negligible. Electron density evolution follows the quasi-neutrality constraint. The plasma pressure $p$ evolves under impurity ionization/recombination and radiative cooling effects, exhibiting pronounced sensitivity to the impurity-induced modifications of local density and temperature profiles. The initial impurity source is neutral gas and localized near the plasma separatrix, with its quantity increasing linearly over $0.1$ ms and featuring Gaussian distributions in both poloidal and toroidal directions (Fig. \ref{fig: eq profiles}a).

All particle species are assumed to share identical fluid velocity $\bi{V}$, and the plasma is initially stationary ($\bi{V}_0=0$ m/s). Particle diffusivity $D=1$ m$^2$/s and viscosity $\nu=1\times10^3$ m$^2$/s are adopted for numerical stability. Thermal conductivities are fixed at $\kappa_{\perp}=1$ m$^2$/s and $\kappa_{\parallel}=1\times10^{10}$ m$^2$/s. The perpendicular thermal conductivity $\kappa_{\perp}$ matches the order of experimental values ($\sim1$ m$^2$/s), whereas the particle diffusivity is one order of magnitude larger than typical experimental values ($\sim0.1$ m$^2$/s) \cite{Groebner_2023}. Plasma resistivity $\eta$ follows the Spitzer model: $\eta\propto Z_{\rm eff} T_e^{-3/2}$, where $Z_{\rm eff}=1+\left(\sum n_Z/n_e\right) \left(\left\langle Z\right\rangle^2-\left\langle Z\right\rangle\right), \left\langle Z\right\rangle=\left(n_e-n_i\right)/\sum n_Z$ represents the effective charge number incorporating impurity ionization/recombination effects.
The numerical implementation employs $40\times80$ bi-quadratic Lagrange polynomial finite elements in the poloidal plane and $22$ Fourier modes spanning toroidal mode numbers $n=0$--$21$. The simulation grid is radially non-uniform, with enhanced resolution in the pedestal region.
A vacuum region surrounds the plasma, and a perfectly conducting wall boundary condition is imposed at the edge of simulation domain.

\subsection{Equilibrium}
\begin{figure}[ht]
	\begin{center}
		\includegraphics[width=0.54\linewidth]{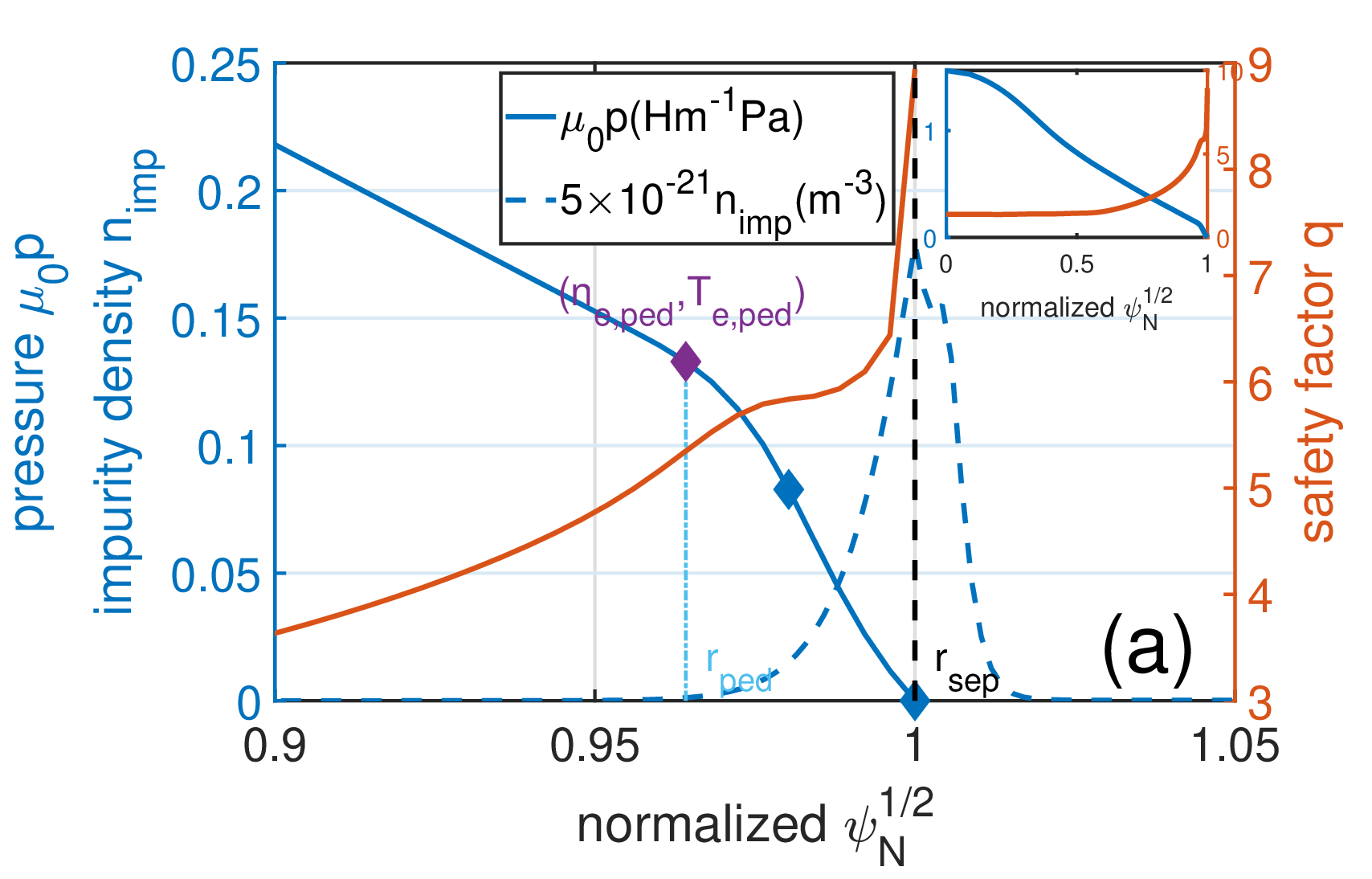}
		\includegraphics[width=0.45\linewidth]{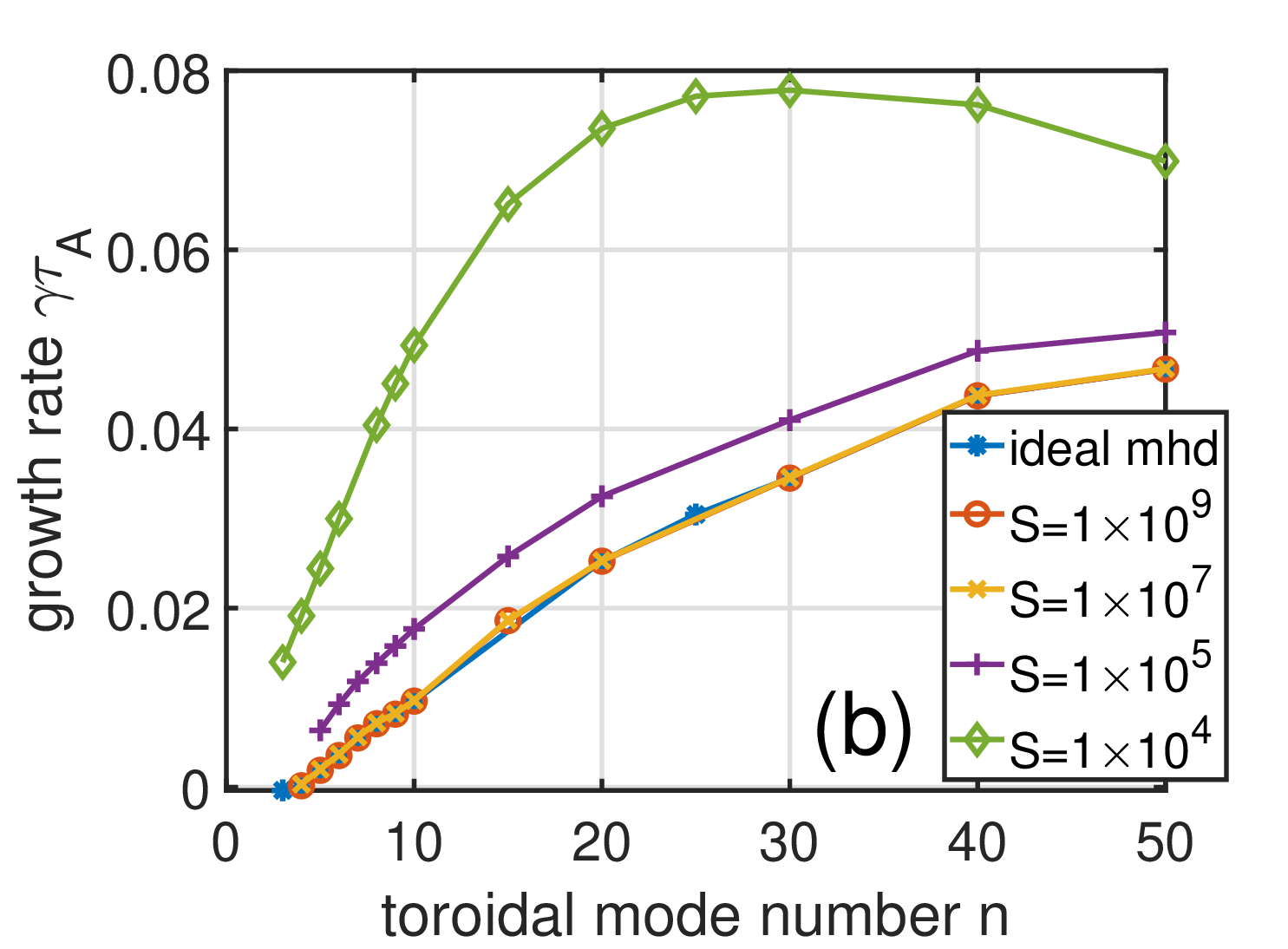}
	\end{center}
	\caption{(a) Equilibrium profiles of plasma pressure $\mu_0p$ (blue solid line) and safety factor $q$ (orange solid line) in the pedestal region, with the plasma separatrix marked using a black dashed line. The locations of pedestal density $n_{\rm e,ped}$ and temperature $T_{\rm e,ped}$ are marked using purple diamonds. The impurity ion density profile is shown as a blue dashed line. The top-right inset displays global profiles using the same color schemes. (b) Linear growth rates versus toroidal mode number $n$ for various Lundquist number $S$ regimes from the linear MHD simulations using NIMROD code.}
	\label{fig: eq profiles}
\end{figure}

The plasma equilibrium employed in the simulations is based on the CFETR hybrid scenario with a lower single-null configuration \cite{Zhuang_2019}. Key parameters are summarized in Tab. \ref{tb:equilibrium parameters}, with equilibrium profiles illustrated in Fig. \ref{fig: eq profiles}(a). The pedestal region exhibits a width of approximately $6$ cm at the outboard mid-plane, featuring pedestal height $n_{\rm e,ped}=6.5\times10^{19}$ m$^{-3}$ and $T_{\rm e,ped}=4.7$ keV, where $n_{\rm e,ped}$ and $T_{\rm e,ped}$ are the pedestal electron density and temperature, respectively. Both the maximum local and flux-surface-averaged pressure gradients are located at the pedestal center. The steep pedestal pressure profile induces substantial current density, resulting in reduced magnetic shear and a nearly flat safety factor profile within the narrow pedestal. The large pedestal pressure gradient drives ballooning mode instabilities, while the high current density stabilizes high-$n$ ballooning modes yet destabilizes low- and intermediate-$n$ kink/peeling mode instabilities.

\begin{table}
	\caption{\label{tb:equilibrium parameters}Key parameters of the equilibrium}
	\footnotesize\rm
	\begin{tabular*}{\textwidth}{@{}l*{15}{@{\extracolsep{0pt plus12pt}}l}}
		\br
		Parameter & Symbol & Value & Unit\\
		\mr
		Minor radius & $a$ & $2.205$ & m\\
		Major radius & $R_0$ & $7.228$ & m\\
		Plasma current & $I_p$ & $1.267\times10^1$ & MA\\
		Toroidal magnetic field & $B_t$ & $6.503$ & T\\
		Central safety factor & $q_0$ & $1.380$ & Dimensionless\\
		Edge safety factor & $q_{95}$ & $6.039$ & Dimensionless\\
		Central electron temperature & $T_{\rm e,core}$ & $3.082\times10^1$ & keV\\
		Central electron density & $n_{\rm e,core}$ & $1.258\times10^{20}$ & m$^{-3}$\\
		Edge electron temperature & $T_{\rm e,edge}$ & $2.862$ & keV\\
		\br
	\end{tabular*}
\end{table}

Linear simulations demonstrate that mode growth rates increase monotonically with toroidal mode number $n$ with both ideal MHD conditions and high Lundquist numbers (Fig. \ref{fig: eq profiles}b). The most unstable high-$n$ modes indicate that ballooning-type instabilities play a dominant role in the ELM crash dynamics for this scenario. Subsequent nonlinear simulations employ high Lundquist numbers ($S\ge1\times10^7$), where notably, the mode growth rates exhibit minimal variation across this regime. In low Lundquist number regimes, elevated plasma resistivity boosts growth rates and shifts down the mode number of the dominant instability.
Plasma resistivity can be enhanced after impurity seeding through radiative cooling and ionization processes, thereby modifying MHD instability growth and the resultant ELM dynamics. However, modifications to the pedestal pressure profile, which primarily arise from the impurity-seeding induced density perturbations, are the focus of this study due to their dominant roles in the nonlinear ELM dynamics (discussed in subsequent sections). This dominance likely stems from the equilibrium pedestals that are pressure-gradient-limited, with ballooning-type modes prevailing in the instability spectrum. Whereas plasma resistivity plays a secondary role here, it may gain significance in scenarios where pedestal stability is governed by kink/peeling mode instabilities or in low Lundquist number regimes, thus warranting further investigations.

\section{Nonlinear dynamics of natural ELM crash}
\label{Sec: nonlinear dynamic of ELM crash}

\begin{figure}[ht]
	\begin{center}
		\includegraphics[width=0.55\linewidth]{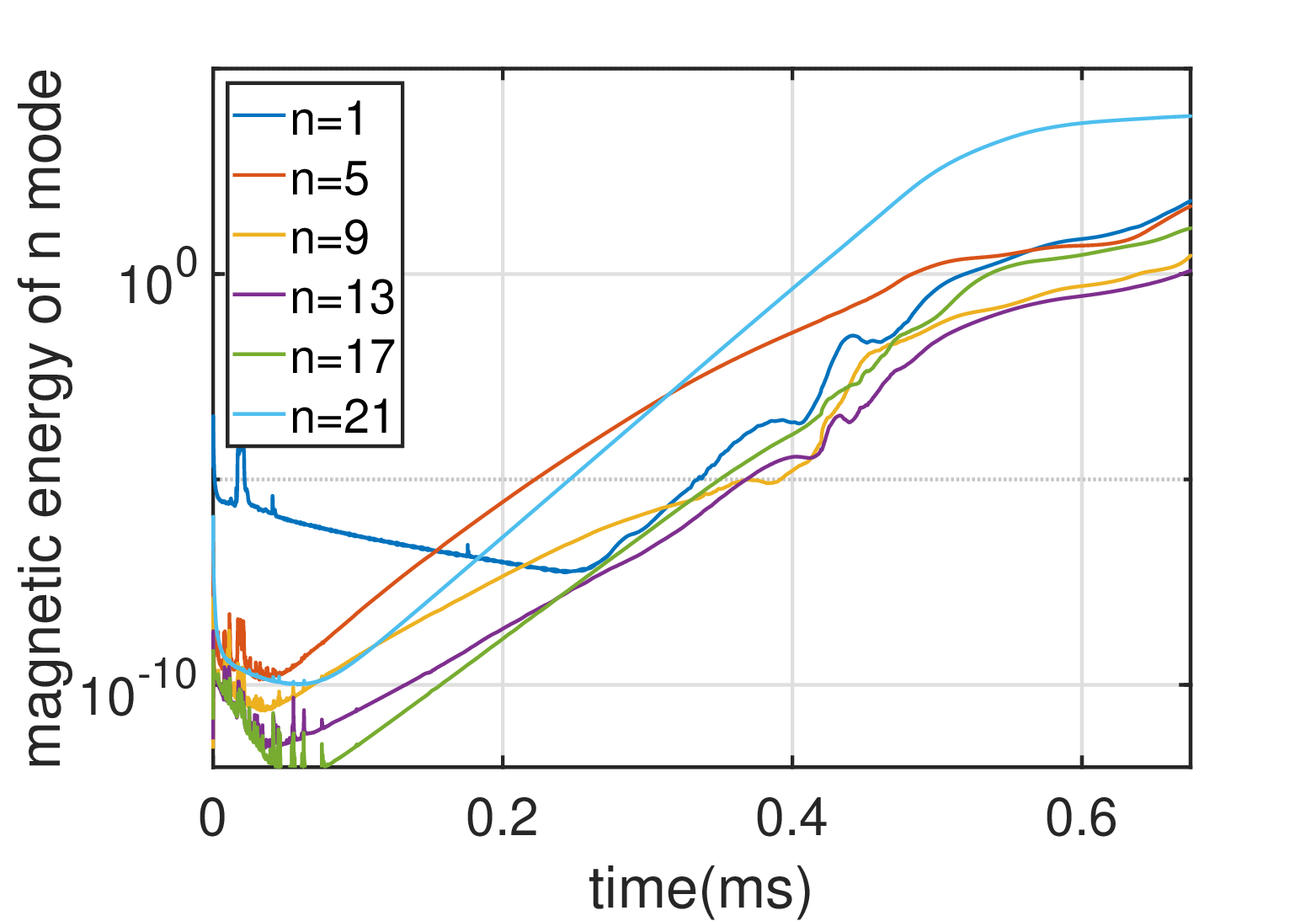}
	\end{center}
	\caption{Magnetic energy evolution of selected toroidal modes $n$ versus time, with the Lundquist number in the plasma core region set to $S=1\times10^7$.}
	\label{fig: elm magnetic nmodes}
\end{figure}

\begin{figure}[ht]
	\begin{center}
		\includegraphics[width=0.45\linewidth]{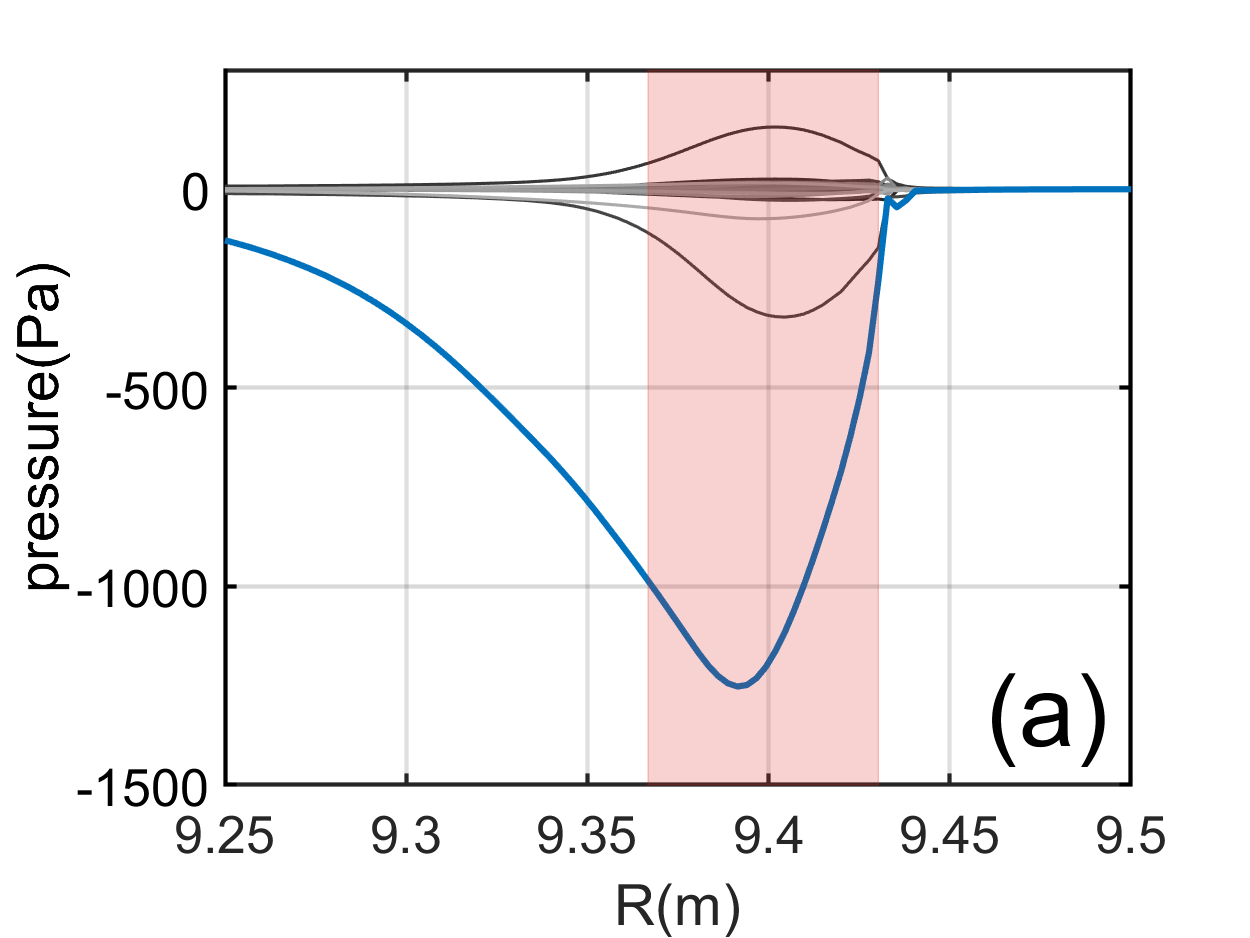}
		\includegraphics[width=0.26\linewidth]{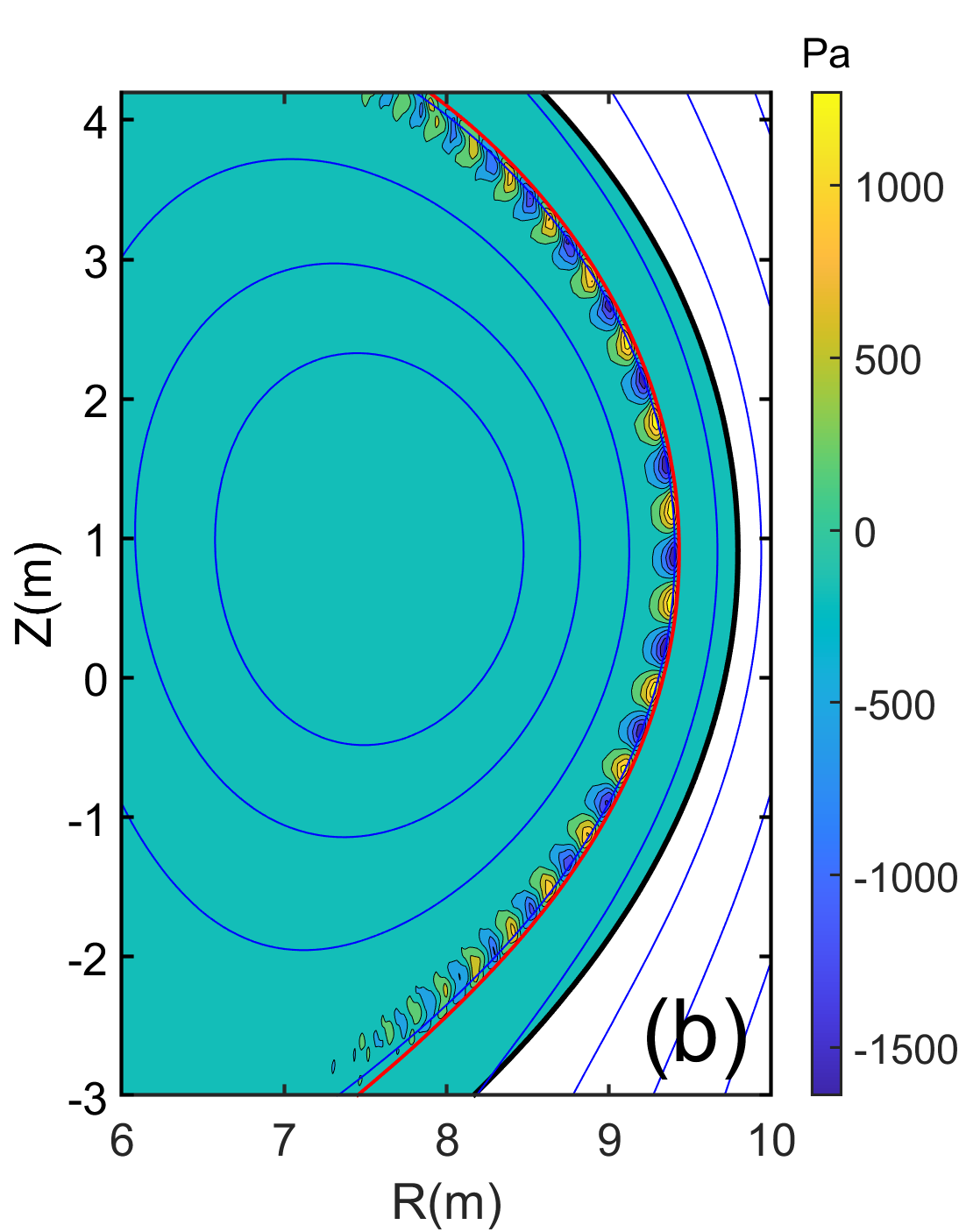}
	\end{center}
	\caption{(a) Pressure profiles of selected toroidal mode components at $t=0.45$ ms, with the $n=21$ mode (blue line) and the pedestal region (red-shaded area) highlighted. (b) Pressure contour of the toroidal mode $n=21$ at $t=0.45$ ms, along with the equilibrium flux surfaces (blue contours) and the plasma separatrix (red line).}
	\label{fig: elm high n mode structure}
\end{figure}

\begin{figure}[ht]
	\begin{center}
		\includegraphics[width=1.0\linewidth]{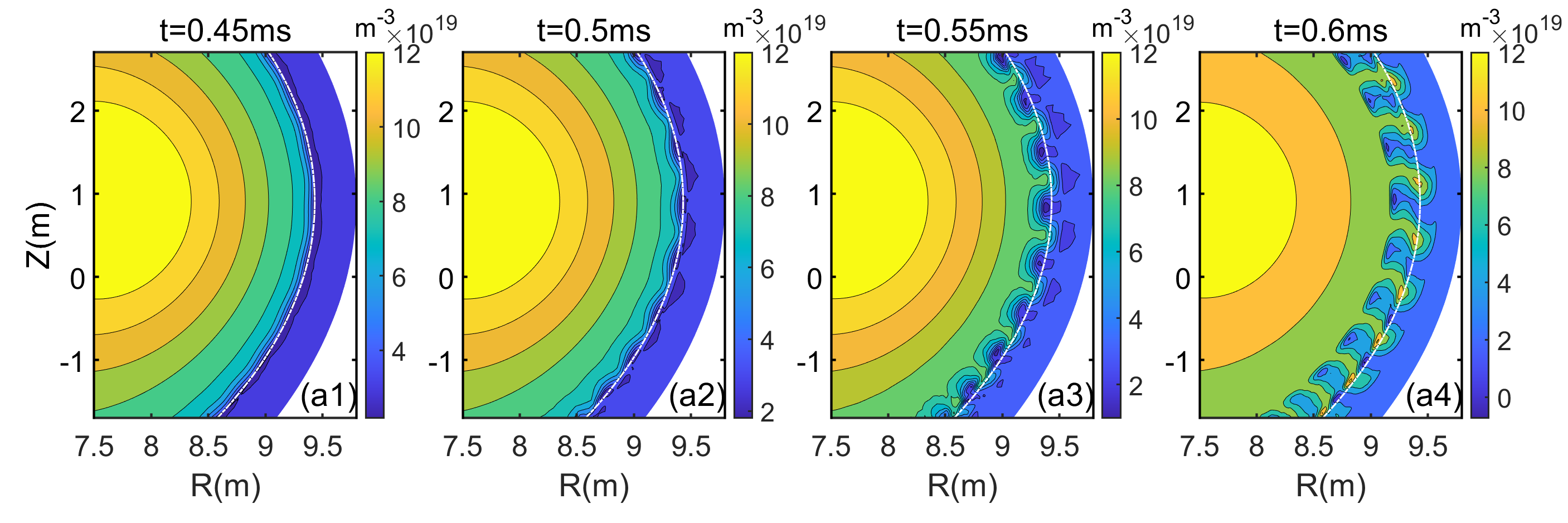}
		\includegraphics[width=0.45\linewidth]{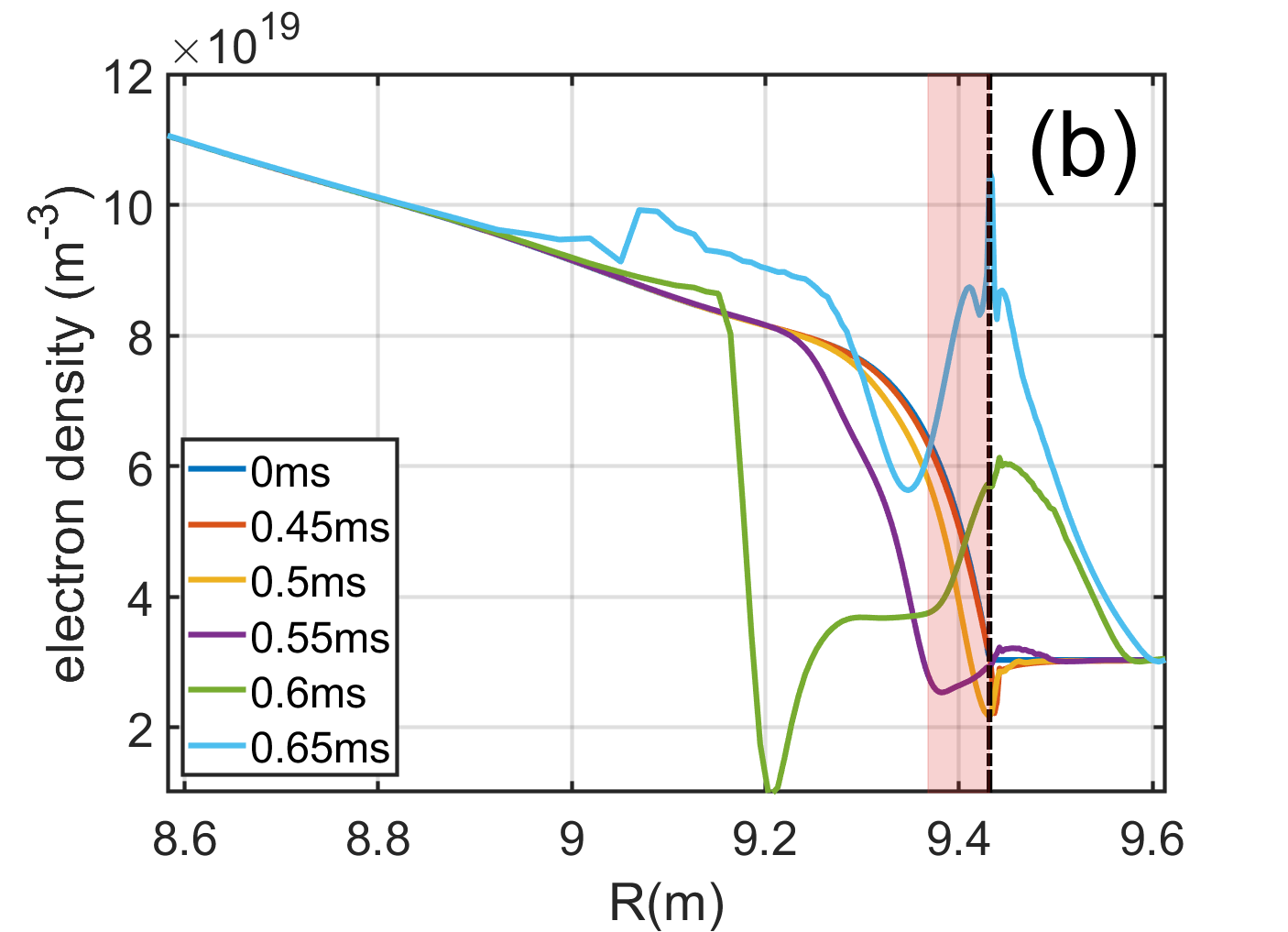}
		\includegraphics[width=0.45\linewidth]{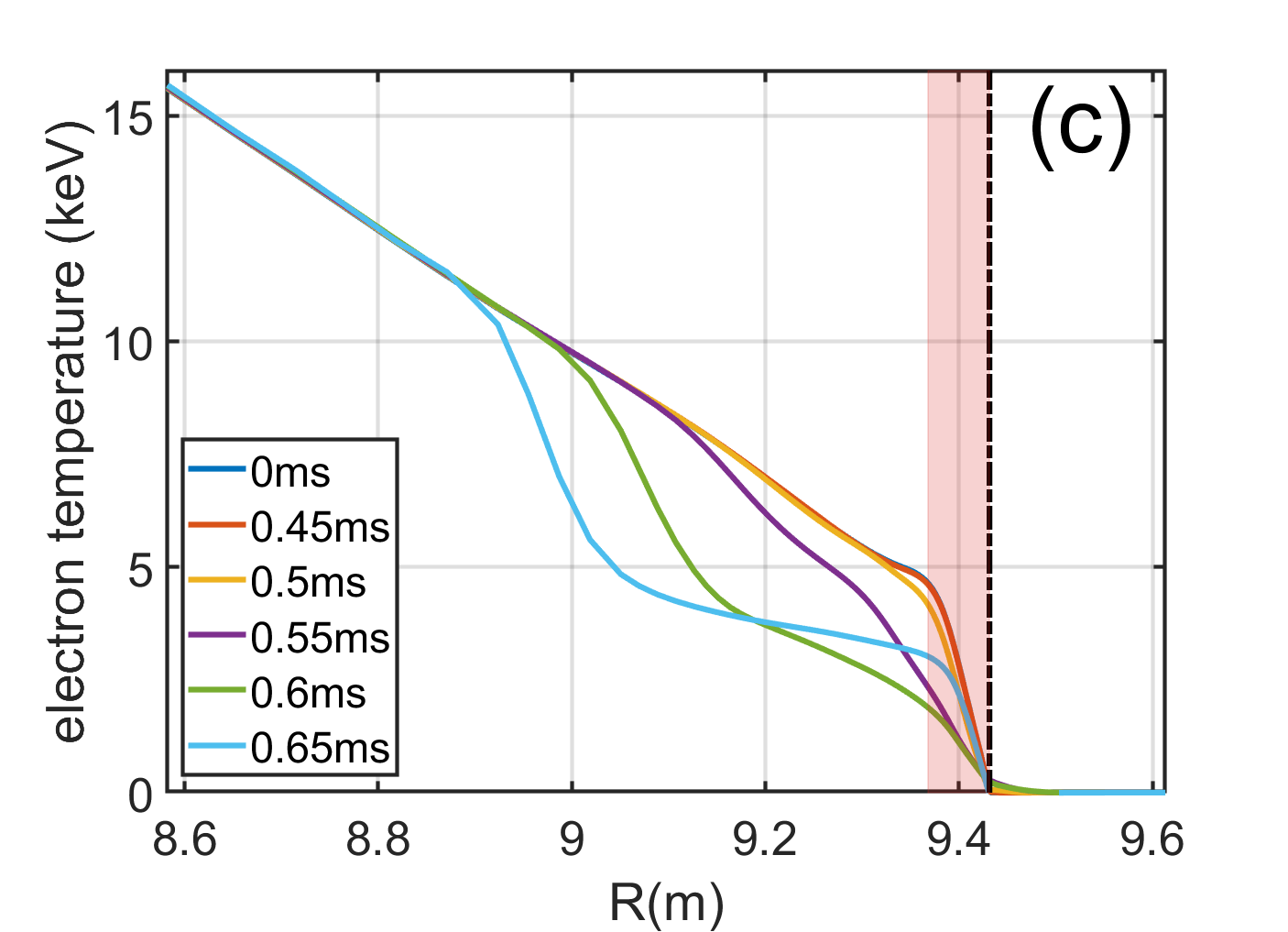}
	\end{center}
	\caption{(a1-a4) Electron density contours at $t=0.45$ ms, $0.5$ ms, $0.55$ ms, and $0.6$ ms, with the plasma separatrix marked using white dashed lines. (b) Outboard mid-plane radial electron density and (c) temperature profiles at a sequence of time. The plasma pedestal region (red-shaded area) and separatrix (black dashed) are highlighted.}
	\label{fig: elm filaments}
\end{figure}

This section demonstrates the dynamics of a natural ELM crash without prior impurity seeding in terms of filamentary structures, mode structures, the correlation between MHD instabilities and ELM triggering, as well as pedestal collapse and energy loss during the ELM crash.
The modes grow from minimal initial seed perturbations, progressing through three distinct stages as illustrated in Fig. \ref{fig: elm magnetic nmodes}. Initially, the seed perturbations decay upon simulation onset, with the $n=1$ mode undergoing an extended decay period. Subsequently, the modes transition into a rapid exponential growth phase, during which the $n=21$ mode becomes dominant. In the final stage, the modes approach saturation.
The growth of the $n=21$ mode correlates closely with the dynamics of the ELM crash---specifically its saturation time and amplitude. This relationship will be discussed in detail in subsequent sections.
All modes remain localized in the pedestal region, with the $n=21$ mode's peak positioned near the pedestal waist region at the ELM crash onset (Fig. \ref{fig: elm high n mode structure}a). Furthermore, the $n=21$ mode exclusively localizes on the outboard bad-curvature side, in agreement with the ballooning mode instability characteristics (Fig. \ref{fig: elm high n mode structure}b).
The ELM crash and pedestal collapse follow the exponential mode growth during the second stage, manifested by filamentary structure development shown in Figs. \ref{fig: elm filaments}(a1-a4). The filamentary structures' spatial patterns resemble those of ballooning-type instabilities, both localized on the bad-curvature side. The pedestal collapse commences at $t=0.45$ ms, initiating at the pedestal top. The density profile collapses and extends beyond the plasma separatrix due to ballooning-interchange convective cells on the bad-curvature side (Fig. \ref{fig: elm filaments}b). In contrast, the temperature profile collapses inward with negligible separatrix extension (Fig. \ref{fig: elm filaments}c), a result attributable to the rapid parallel thermal transport along open field lines outside the plasma separatrix. After $t=0.55$ ms, the collapse of plasma profile propagates inward following the complete pedestal destruction. In absence of external heating or particle fueling, plasma profiles undergo irreversible collapse once the ELM crash begins. However, the outward fluxes released from the central plasma profile collapse partially restore the pedestal profiles, which is evident in the $t=0.65$ ms profiles in Figs. \ref{fig: elm filaments}(b) and (c). The general ELM dynamics outlineed above agrees well with experiments. For example, the characteristics of the filamentary structure are consistent with MAST experimental observations \cite{Kirk_PhysRevLett.96.185001,Kirk_2007}. Additionally, precursors with high-$n$ mode number are widely observed to be localized in the pedestal region and grow up before ELM onset \cite{Kirk_2014}. On KSTAR, ELM evolution is also characterized in three stages: initial linear phase, quasi-steady state, and crash phase \cite{Yun_PhysRevLett.107.045004}.

\begin{figure}[ht]
	\begin{center}
		\includegraphics[width=0.55\linewidth]{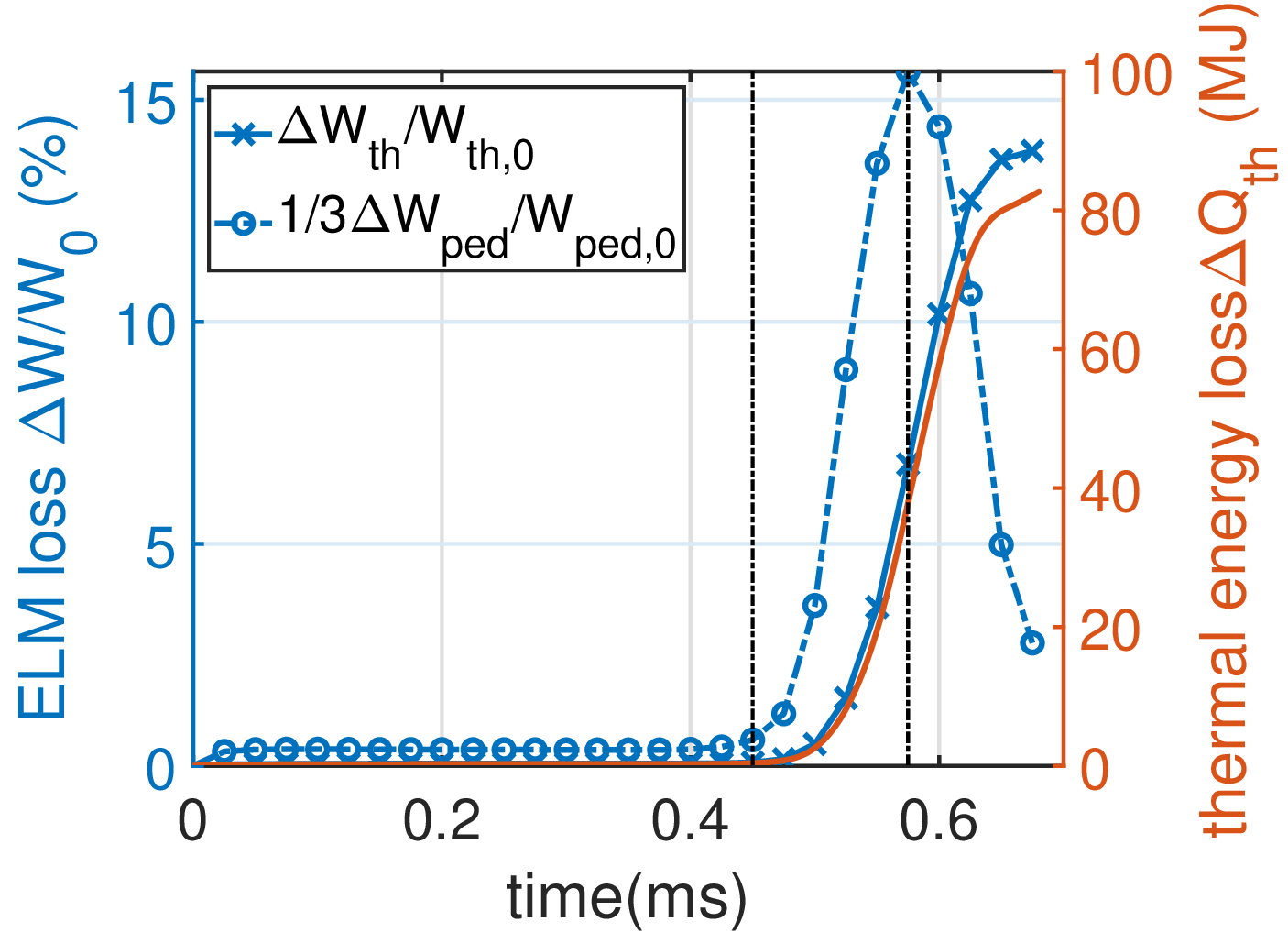}
	\end{center}
	\caption{ELM energy losses (blue lines) and the total plasma thermal energy loss $\Delta Q_{th}$ (orange solid line) as functions of time. The total ELM loss $\Delta W_{th}/\Delta W_{th,0}$ is defined by $W_{th}=\int3n_eT_e \rm dS$ (integrated over the entire plasma region), where $W_{th,0}$ denotes the initial total thermal energy and $\Delta W_{th}=W_{th,0}-W_{th}$. The pedestal ELM loss $\Delta W_{ped}/\Delta W_{ped,0}$ is calculated via $W_{ped}=\int_{0}^{2\pi} {\rm d}\theta \int_{r_{ped}}^{r_{sep}} 3n_eT_e \rm dr$, limited to the pedestal region between the pedestal top $r_{ped}$ and the plasma separatrix $r_{sep}$, where $W_{ped,0}$ represents the initial pedestal thermal energy and $\Delta W_{ped}=W_{ped,0}-W_{ped}$.}
	\label{fig: elm size}
\end{figure}

To quantify ELM energy loss, Fig. \ref{fig: elm size} compares the ELM losses calculated using two alternative definitions. The pedestal ELM loss $\Delta W_{ped}/W_{ped,0}$ increases rapidly after $t=0.45$ ms, peaking at $t=0.575$ ms upon complete pedestal loss, releasing approximately $50\%$ of the pedestal stored thermal energy. The subsequent decrease results from core plasma fluxes gradually replenishing the pedestal. In contrast, the total ELM loss $\Delta W_{th}/W_{th,0}$ increases monotonically throughout the simulation due to continuous profile collapse, as its calculation encompasses the entire plasma region. For subsequent analysis, the pedestal ELM loss $\Delta W_{ped}/W_{ped,0}$ is adopted as the size of ELM crash, due to its more direct correlation with the pedestal collapse. The rapid increase of the pedestal ELM loss marks the onset of an ELM crash at $t=0.45$ ms, and its peak at $t=0.575$ ms is designated as the end of the crash phase. Notably, the total ELM loss $\Delta W_{th}/W_{th,0}$ reaches approxmiately $6.8\%$ at $t=0.575$ ms, significantly lower than the experimental scaling prediction ($\sim17\%$) \cite{LOARTE2003962}, where the experimental stored energy $W_{th,0}^{exp}$ is estimated using $W_{th,0}^{exp}=3\left( n_{e,ped}T_{e,ped}\right) V_{plasma}$ with pedestal density $n_{e,ped}$ and temperature $T_{e,ped}$. By comparison, in simulations the total stored energy $W_{th,0}$ integrates over entire plasma profiles, implying that the experimental scaling prediction likely overestimates ELM size due to an under-estimated $W_{th,0}^{exp}$ in denominator. Despite this discrepancy, approximately $40$ MJ of thermal energy is released at $t=0.575$ ms, posing severe challenges to PFCs.

\begin{figure}[ht]
	\begin{center}
		\includegraphics[width=0.4\linewidth]{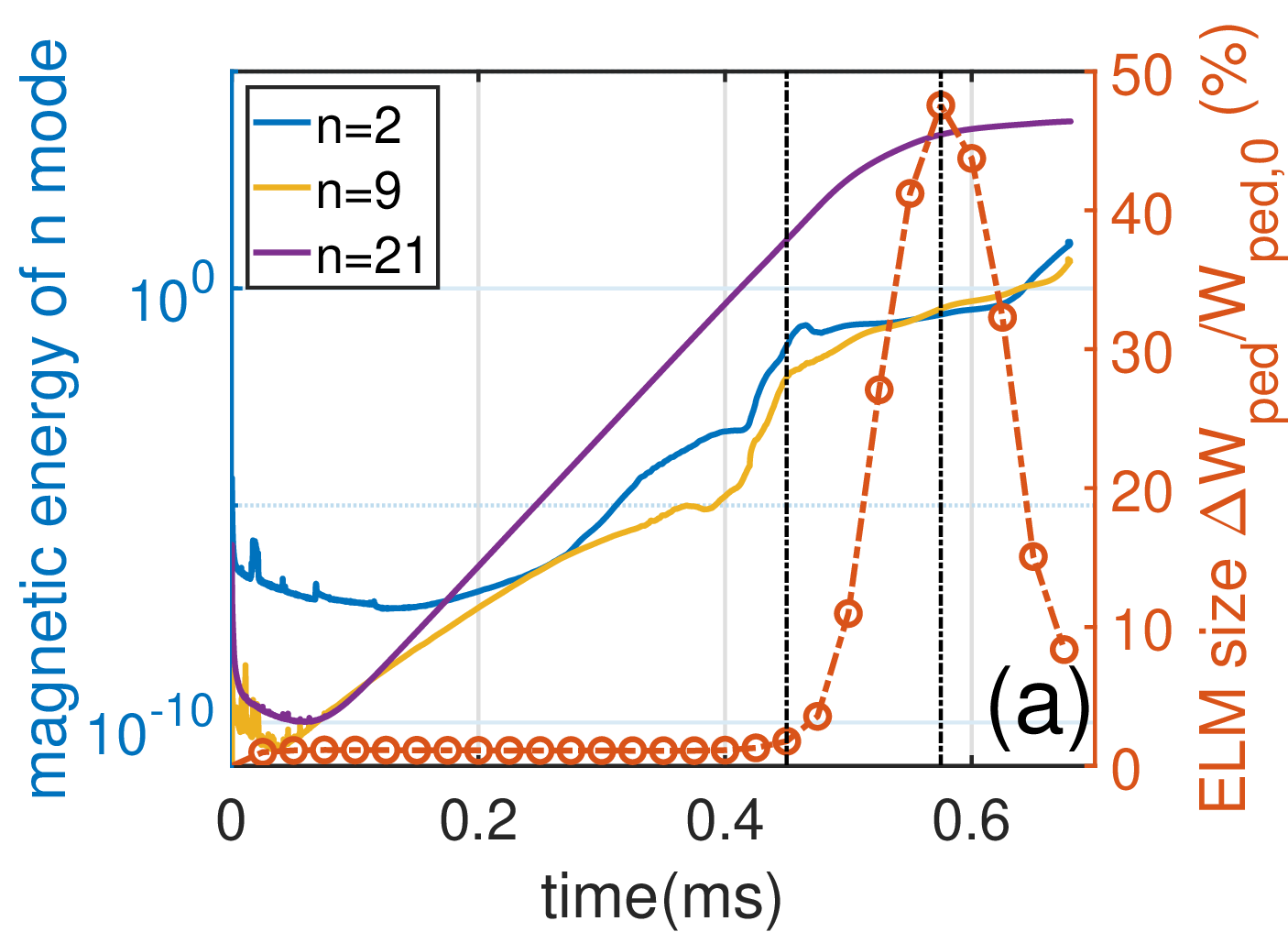}
		\includegraphics[width=0.45\linewidth]{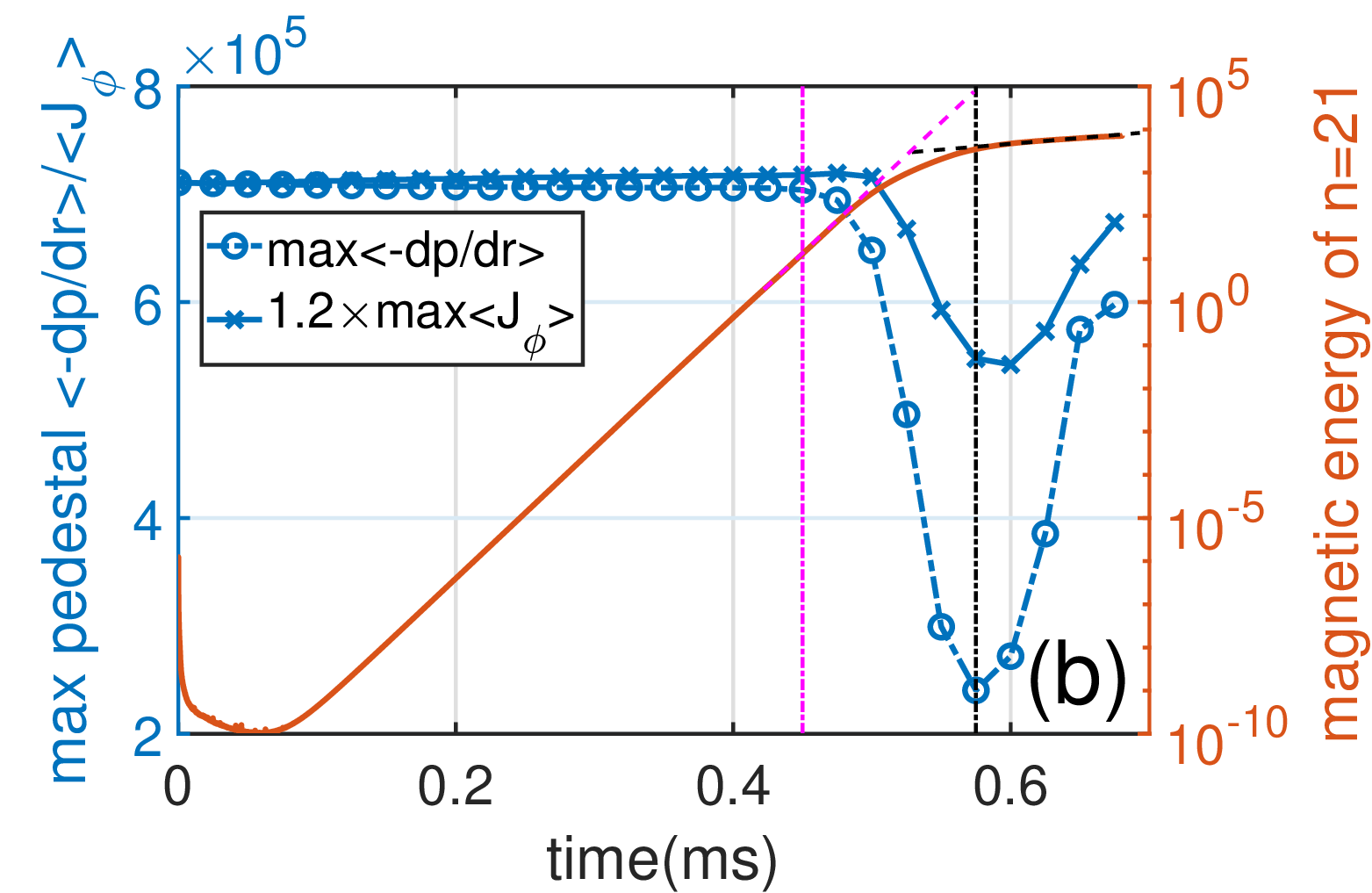}	
	\end{center}
	\caption{(a) Pedestal ELM size (orange dashed line) and magnetic energies of selected toroidal modes are shown as functions of time. (b) Magnetic energy of toroidal mode $n=21$ (orange solid line), maximum flux-surface-averaged pedestal pressure gradient (blue dashed line), and its associated toroidal current density (blue solid line) are plotted as functions of time.}
	\label{fig: elm high n mode growth}
\end{figure}

The $n=21$ mode acts as the primary driver for the ELM crash, while other modes exert secondary influences (Fig. \ref{fig: elm high n mode growth}a). The $n=21$ mode is the most unstable mode at $t=0.45$ ms when rapid pedestal collapse begins, and approaches saturation by the end of ELM crash. Additionally, the growth of low- and intermediate-$n$ modes undergoes a sudden increase at the onset of pedestal collapse ($t=0.45$ ms). The $n=21$ mode, predominantly driven by the pedestal pressure gradient, demonstrates a coupled interaction with the pressure profile evolution (Fig. \ref{fig: elm high n mode growth}b). The steep pressure gradient sustains the exponential mode growth prior to ELM crash. Subsequently, the mode growth rate decelerates and transitions into a slowdown phase concurrently with the decreasing pressure gradient (as indicated by the interval between vertical magenta and black lines in Fig. \ref{fig: elm high n mode growth}b). The mode ultimately enters a secondary steady-growth phase after the pedestal pressure gradient drops to its minimum at $t=0.575$ ms. Moreover, the pedestal current density evolves similarly as the pressure gradient, presumably contributing to the growth of the peeling mode components.
After $t=0.575$ ms, partial pedestal recovery due to the outward flux of core thermal energy elevates both the pressure gradient and current density, driving renewed growth of the $n=21$ mode and other modes rather than sustained saturation.

\begin{figure}[ht]
	\begin{center}
		\includegraphics[width=0.45\linewidth]{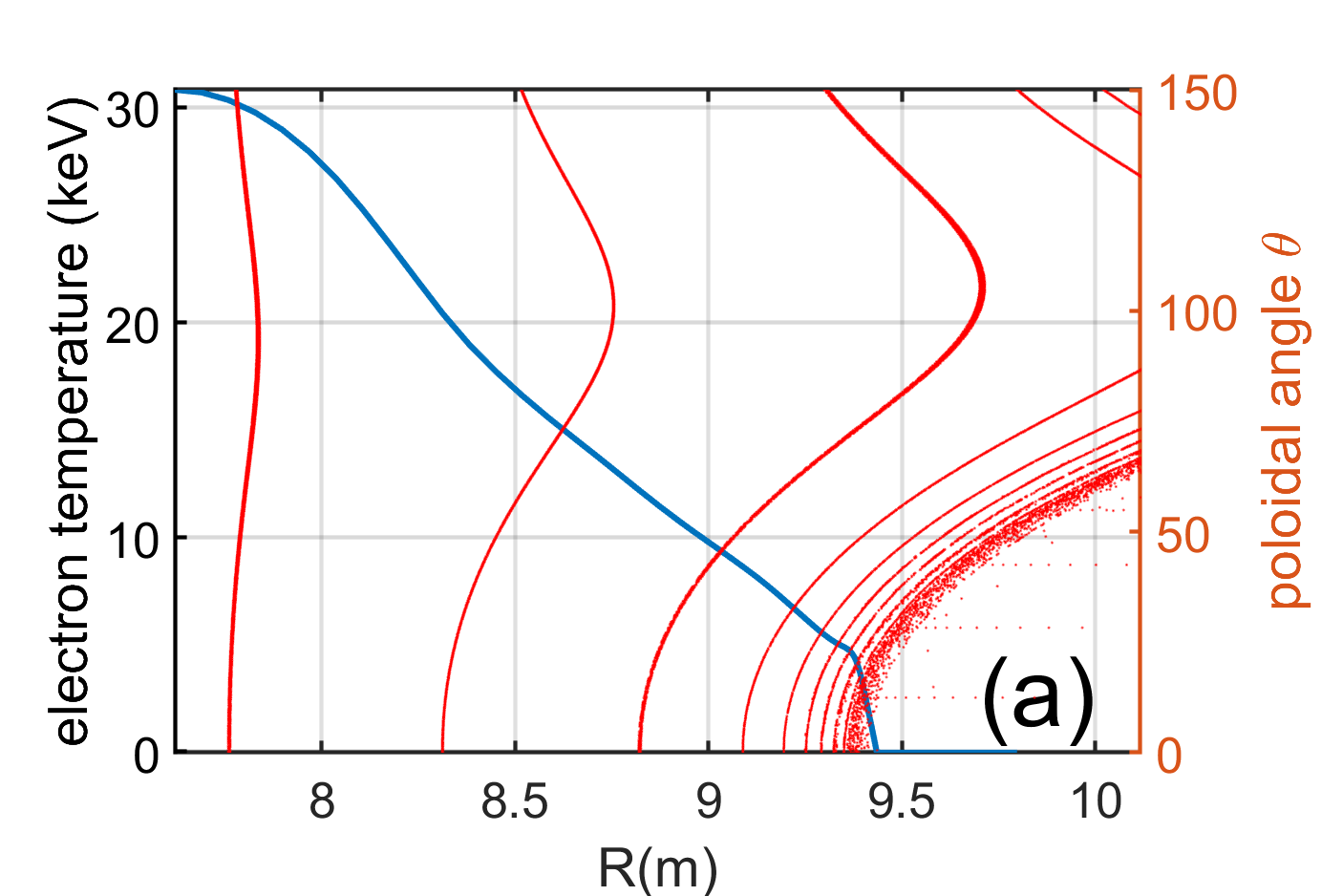}
		\includegraphics[width=0.45\linewidth]{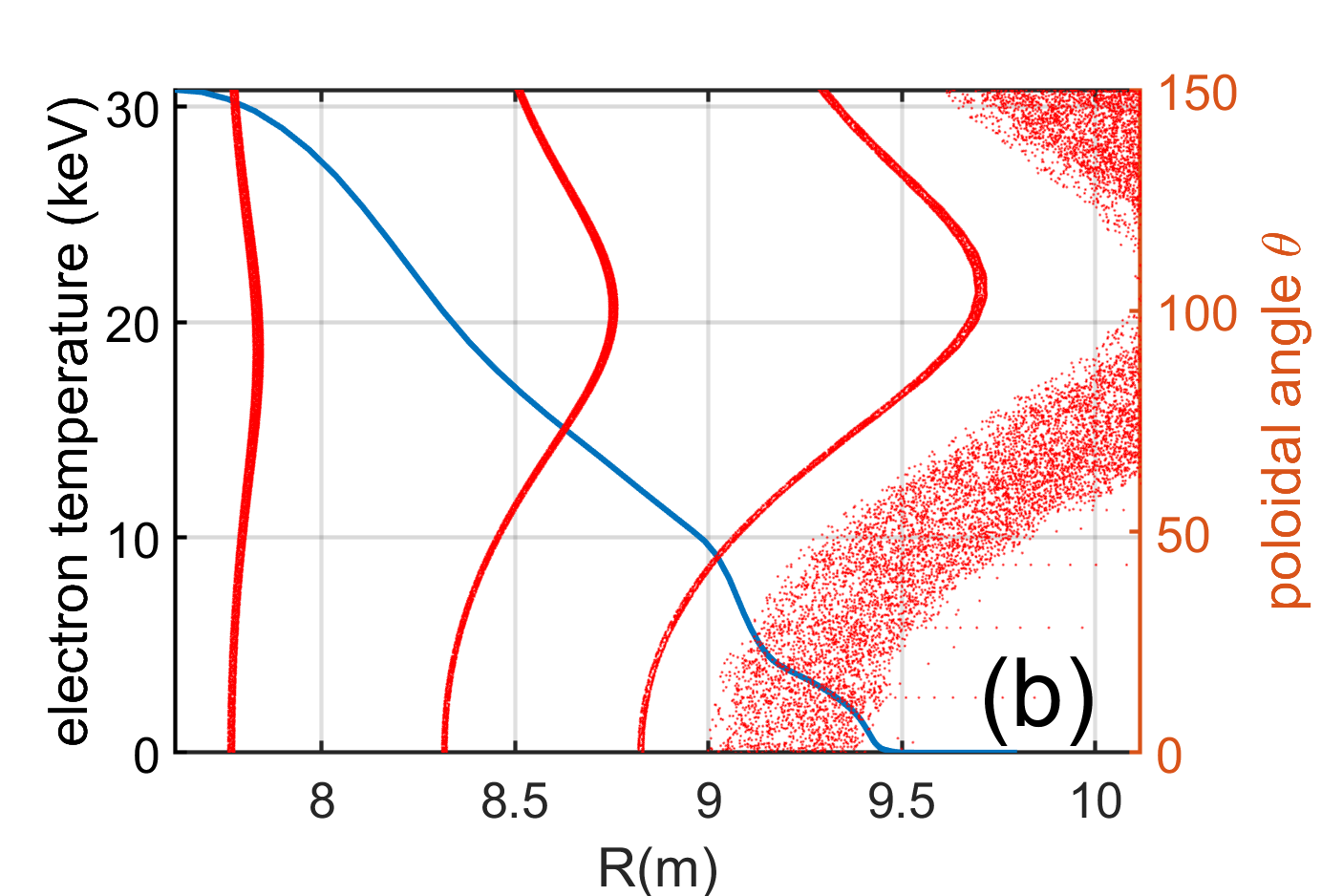}
	\end{center}
	\caption{Poincar\'{e} plots of magnetic field lines (red dots) and temperature profiles (blue lines) at (a) $t=0.45$ ms and (b) $t=0.6$ ms.}
	\label{fig: elm poincare}
\end{figure}
\begin{figure}[ht]
	\begin{center}
		\includegraphics[width=0.45\linewidth]{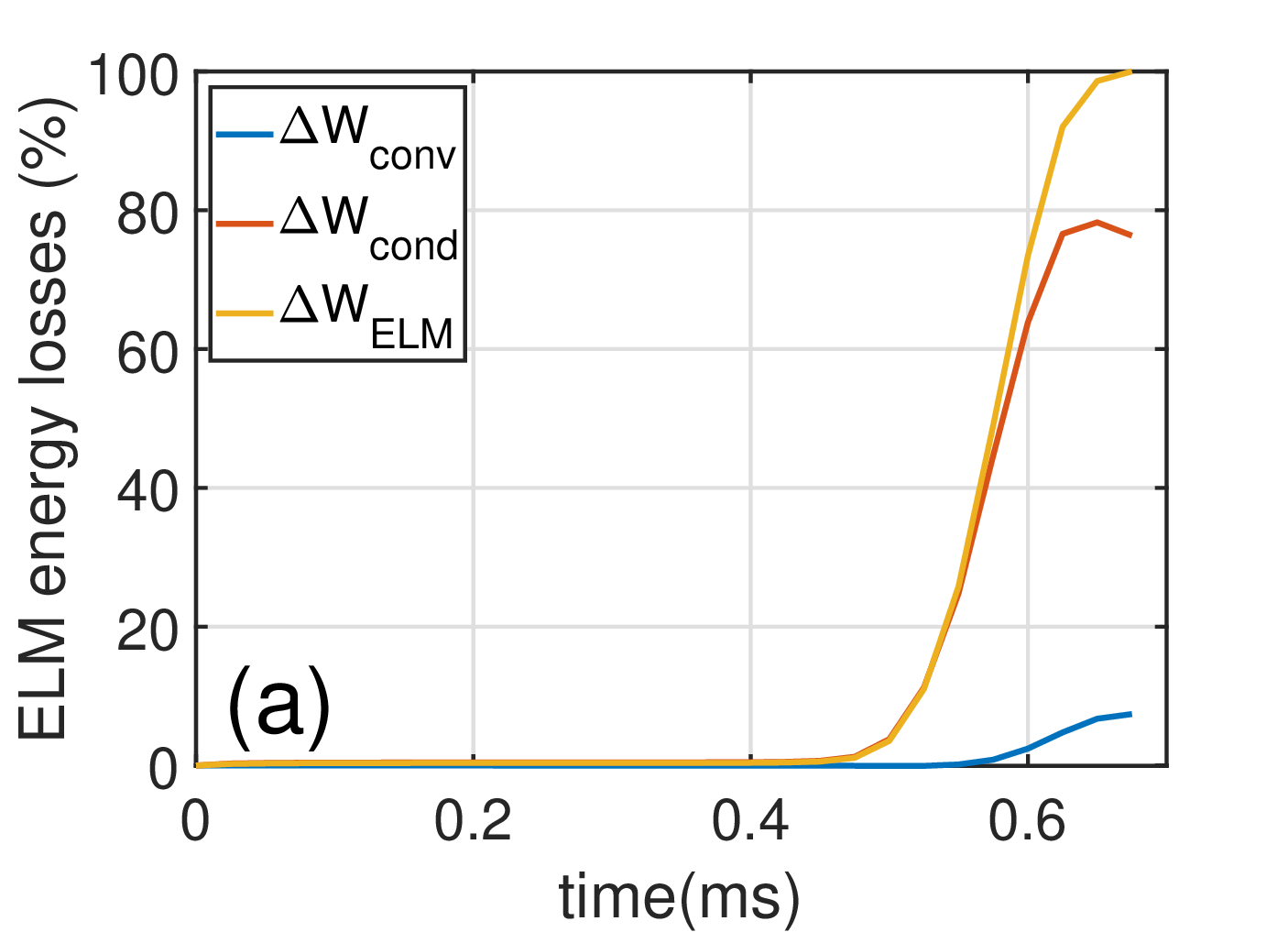}
		\includegraphics[width=0.45\linewidth]{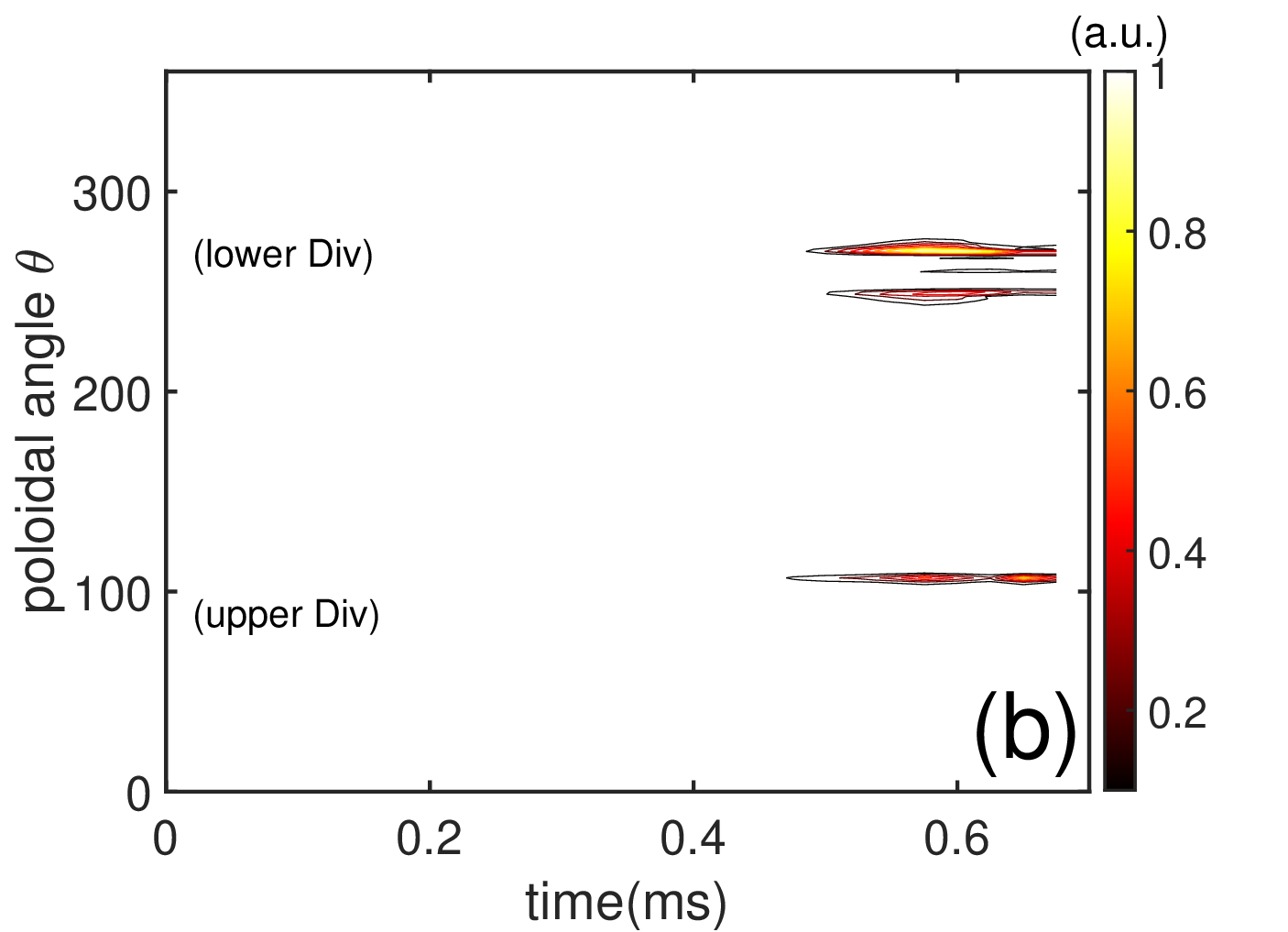}
	\end{center}
	\caption{(a) Definitions of energy losses include: convective loss $\Delta W_{conv}=\int \Delta n\left\langle \Delta T\right\rangle_t\rm dS$, conductive loss $\Delta W_{cond}=\int \left\langle \Delta n\right\rangle_t\Delta T \rm dS$, and total ELM energy loss $\Delta W_{ELM}=\int \Delta n\Delta T \rm dS$, all normalized to $\max(\Delta W_{ELM})$ and shown as functions of time. Here, $\Delta n$ and $\Delta T$ denote instantaneous density and temperature perturbations, respectively, while $\left\langle \Delta n\right\rangle_t$ and $\left\langle \Delta T\right\rangle_t$ represent their time-averaged counterparts. (b) Normalized heat load distribution on the wall along the poloidal direction as a function of time.}
	\label{fig: elm transport}
\end{figure}

During the ELM crash, the magnetic field lines in the pedestal region become stochastic and spread inward, whereas the plasma profiles collapse simultaneously (Figs. \ref{fig: elm poincare}). The energy loss dominated by temperature change $\Delta T_e$ accounts for the majority of total ELM energy loss evolution (Fig. \ref{fig: elm transport}a). This dominance is likely due to rapid thermal transport facilitated by the stochastic magnetic field lines. Released heat flux reaches the wall, resulting in a heat load distribution concentrated at both lower and upper divertors, with peak intensity at the lower divertor (Fig. \ref{fig: elm transport}b). However, the simulation results provide only preliminary insights into ELM crash energy loss. A comprehensive understanding of this process requires implementing advanced thermal transport models and incorporating neutral dynamics in the cold, dense scrap-off layer, which both lie beyond the MHD model intended for the limited scope of this study.

\section{The influences of impurity seeding on ELM dynamics}
\label{Sec: Impurity seeding}

This section examines the nonlinear ELM crash process under various impurity seeding scenarios. Compared to the natural ELM crash scenario, impurity seeding can have substantial impacts on the ELM dynamics. Two key control parameters, specifically the impurity density level and the poloidal seeding location, are systematically studied here.

\subsection{The density level of impurity seeding}
\label{subsection: Impurity density level}

\begin{figure}[ht]
	\begin{center}
		\includegraphics[width=0.55\linewidth]{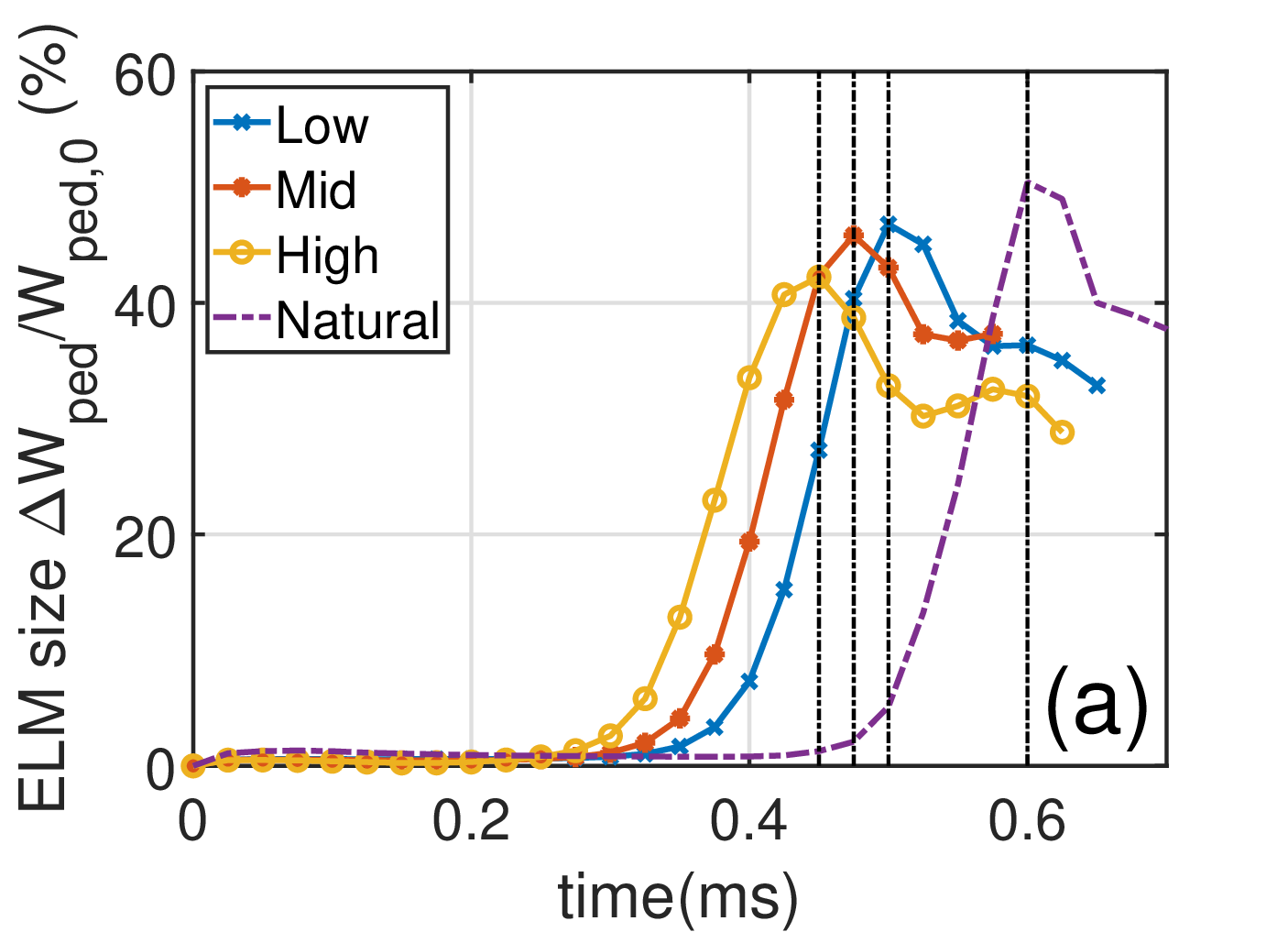}
		\includegraphics[width=0.45\linewidth]{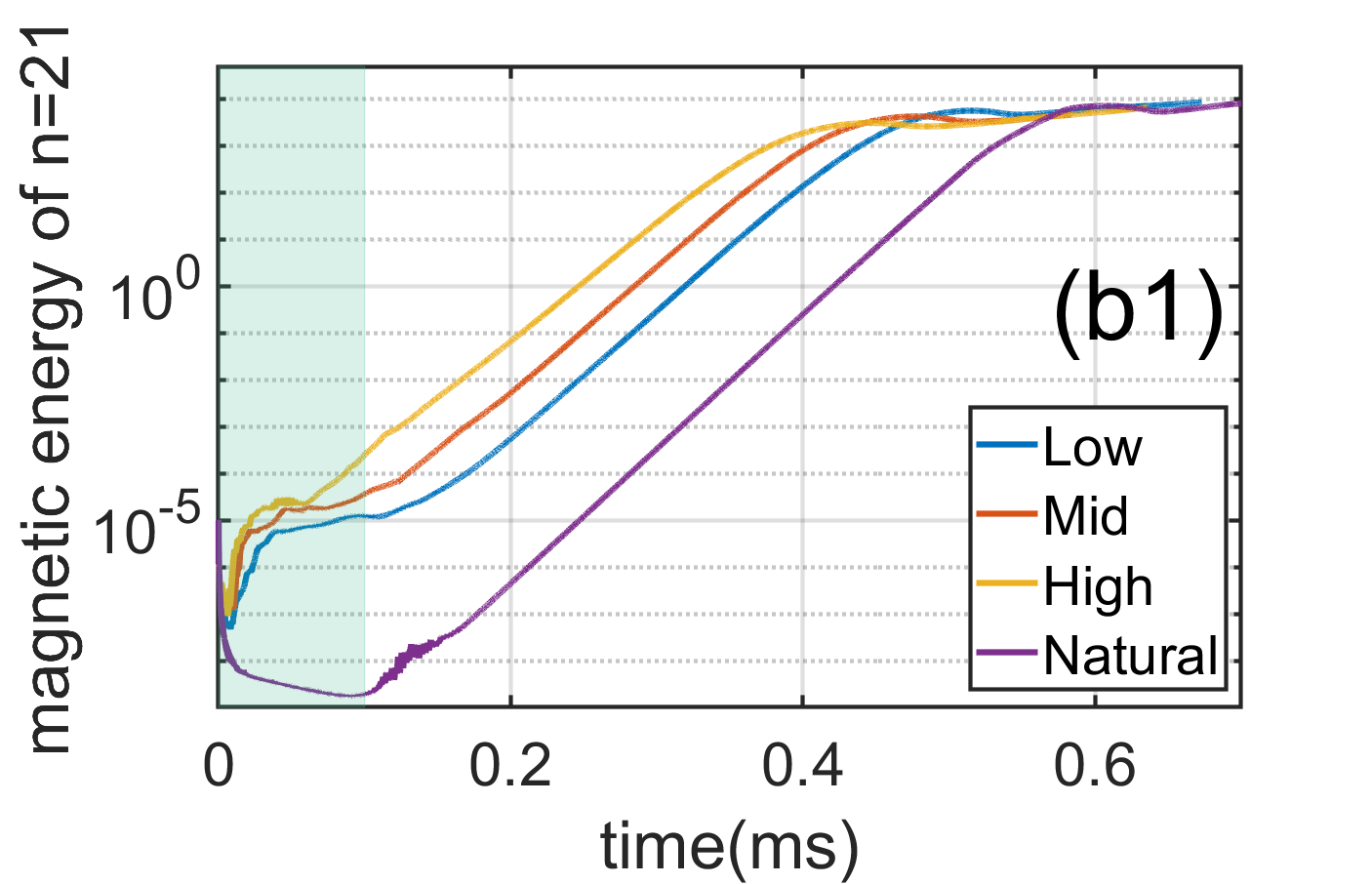}
		\includegraphics[width=0.45\linewidth]{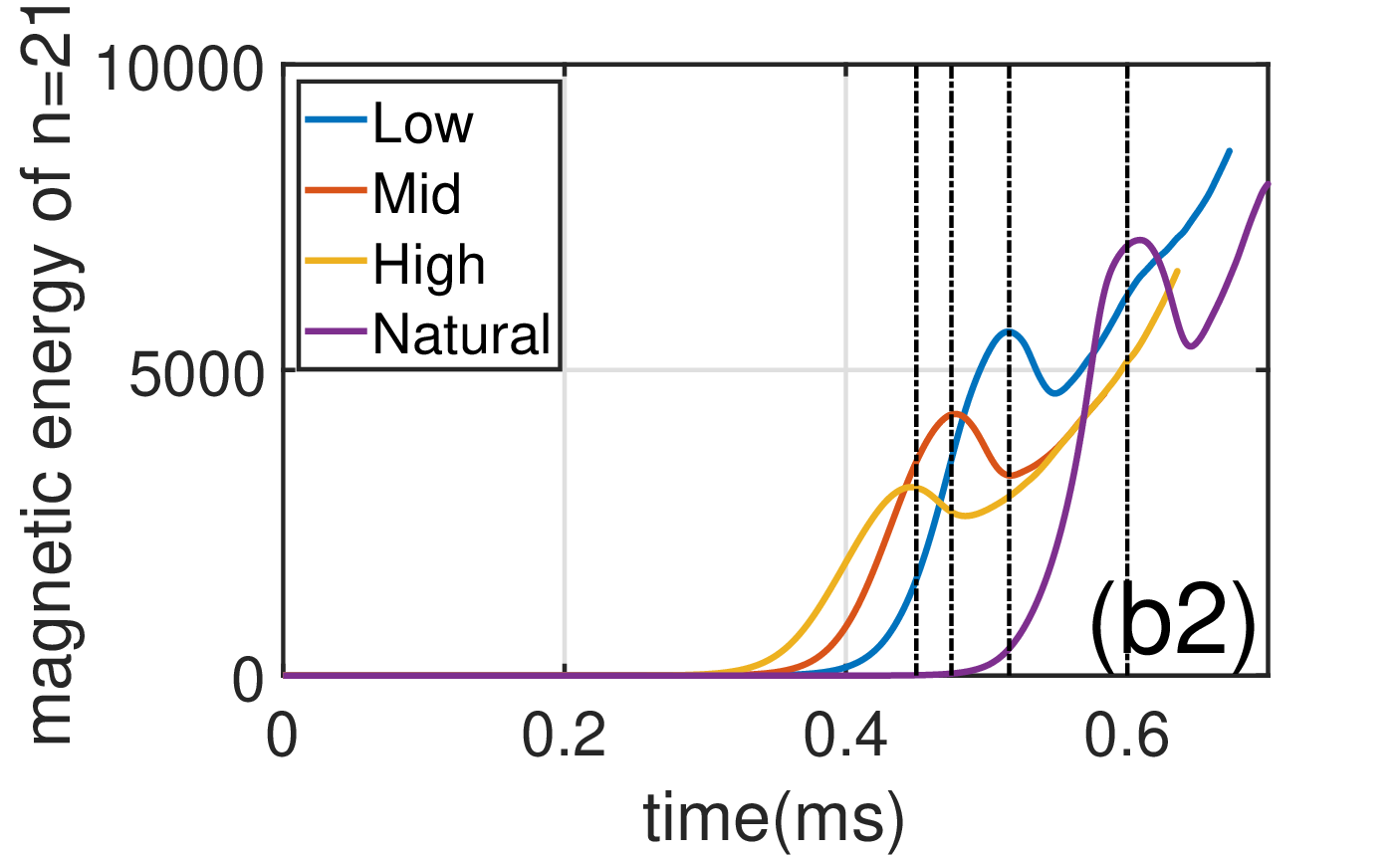}
	\end{center}
	\caption{(a) ELM size and (b1-b2) magnetic energy of the $n=21$ mode are shown as functions of time for various impurity density levels. Cases labeled `Low', `Mid', and `High' correspond to maximum injected impurity ion densities of approximately $1\%$, $5\%$, and $10\%$ of the line-averaged plasma density at ELM crash, respectively. `Natural' denotes the reference case without impurity seeding. The Lundquist number in the plasma core is set to $S=1\times10^9$.}
	\label{fig: impurity resistivity-level lem size}
\end{figure}

Fig. \ref{fig: impurity resistivity-level lem size}(a) shows ELM crashes under various impurity density levels, with impurities seeded at the outboard mid-plane. Compared to the reference case without impurity seeding, impurity seeding with higher impurity density levels accelerates ELM crash onset and reduces the subsequent ELM size. The $n=21$ mode remains the most unstable toroidal component across all cases (Fig. \ref{fig: impurity resistivity-level lem size}b1). However, during the early stage it rapidly grows in impurity seeding cases but decays in the reference case. The ELM size $\Delta W_{ped}/\Delta W_{ped,0}$ reaches its peak simultaneously with that of the $n=21$ mode's growth, and the maximum ELM size scales with the peak amplitude of the mode (Fig. \ref{fig: impurity resistivity-level lem size}b2). After the peak, the growth of the $n=21$ mode is sustained by partial pedestal recovery.
Additionally, impurity radiation cooling contributes negligibly to the total energy loss. For instance, in the `High' impurity density level case, radiation losses account for only approximately $1.5\%$ ($0.7$ MJ) of the $46$ MJ total thermal energy loss.

\begin{figure}[ht]
	\begin{center}
		\includegraphics[width=0.55\linewidth]{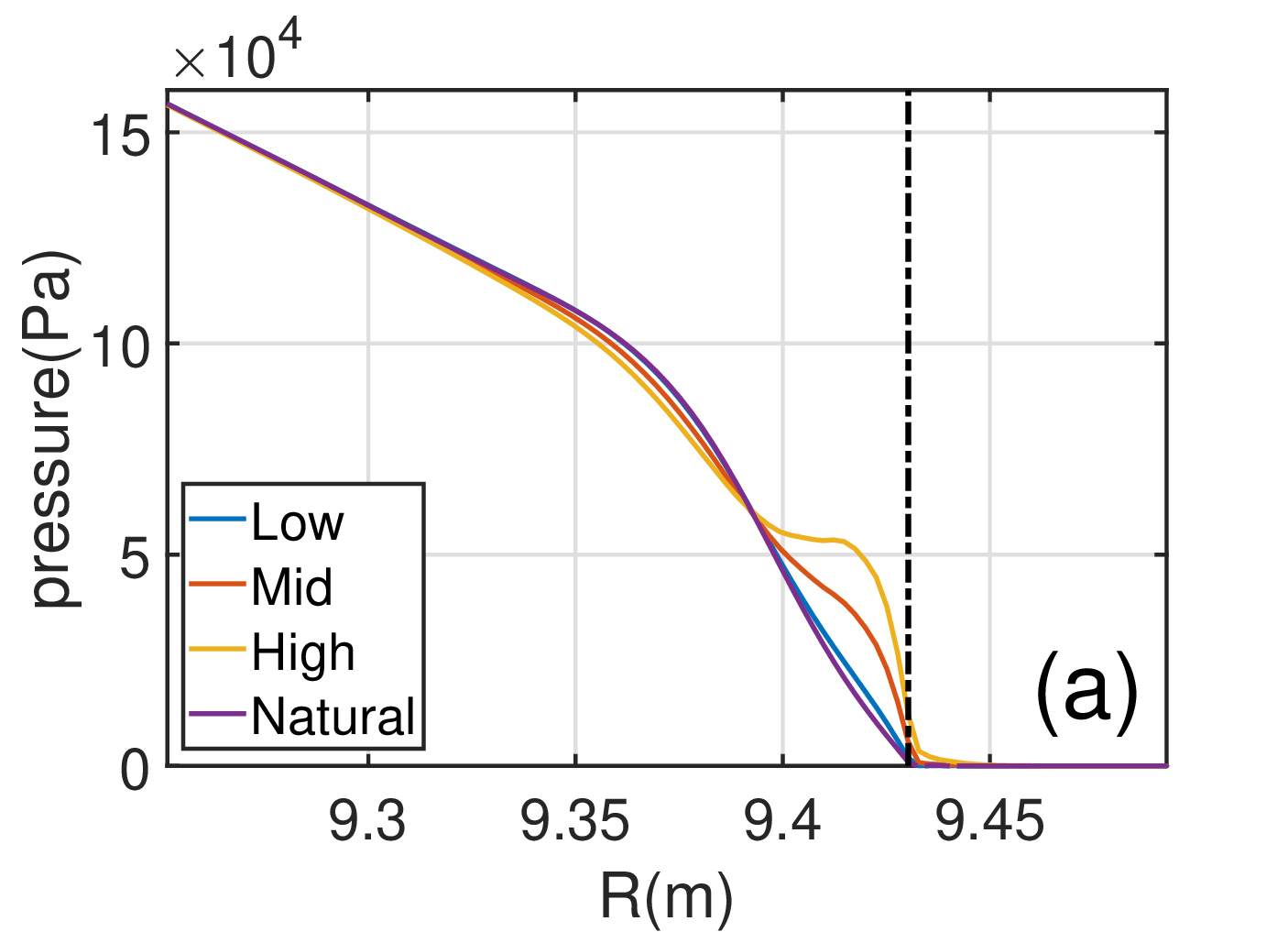}
		\includegraphics[width=0.45\linewidth]{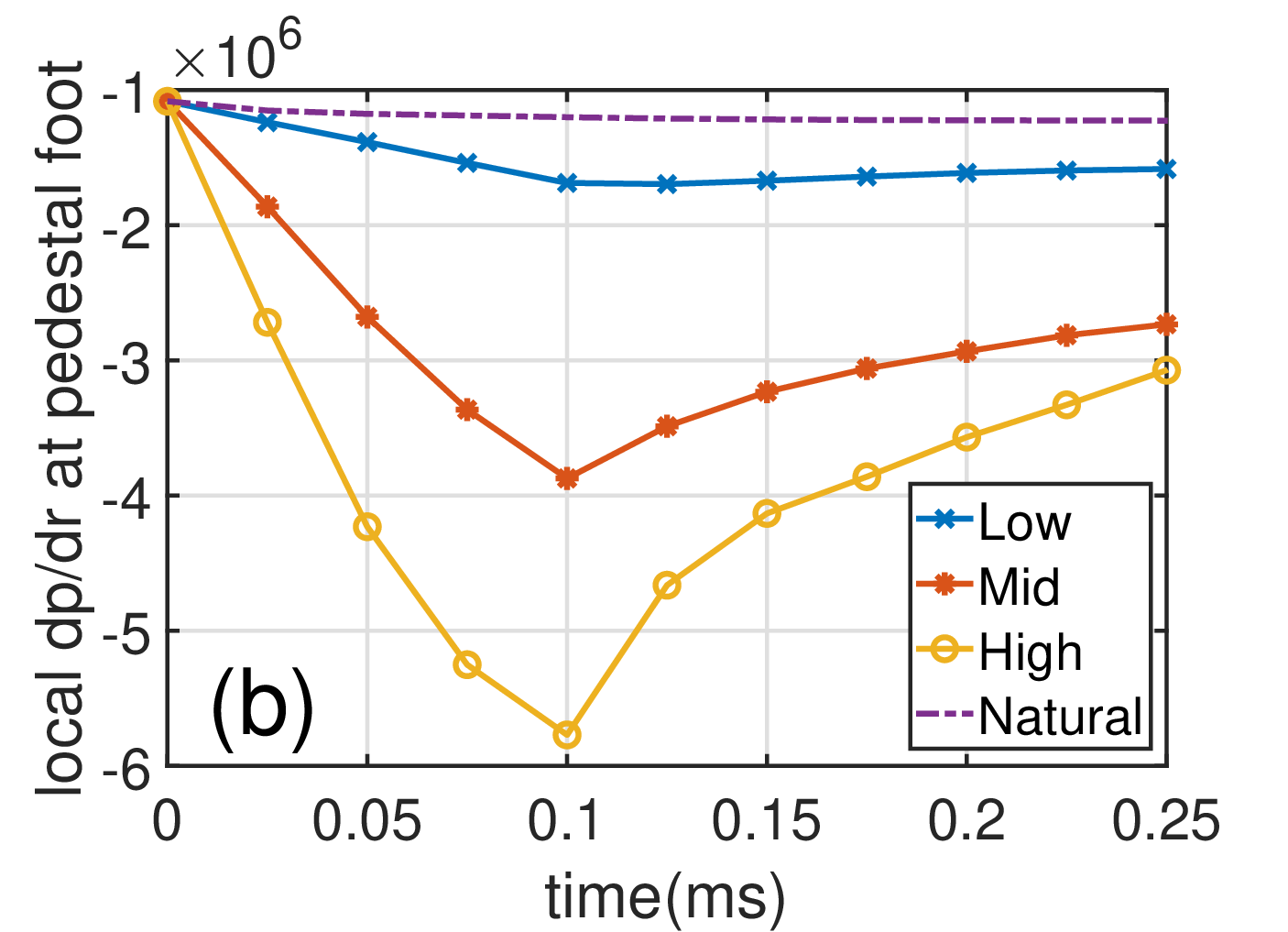}
		\includegraphics[width=0.45\linewidth]{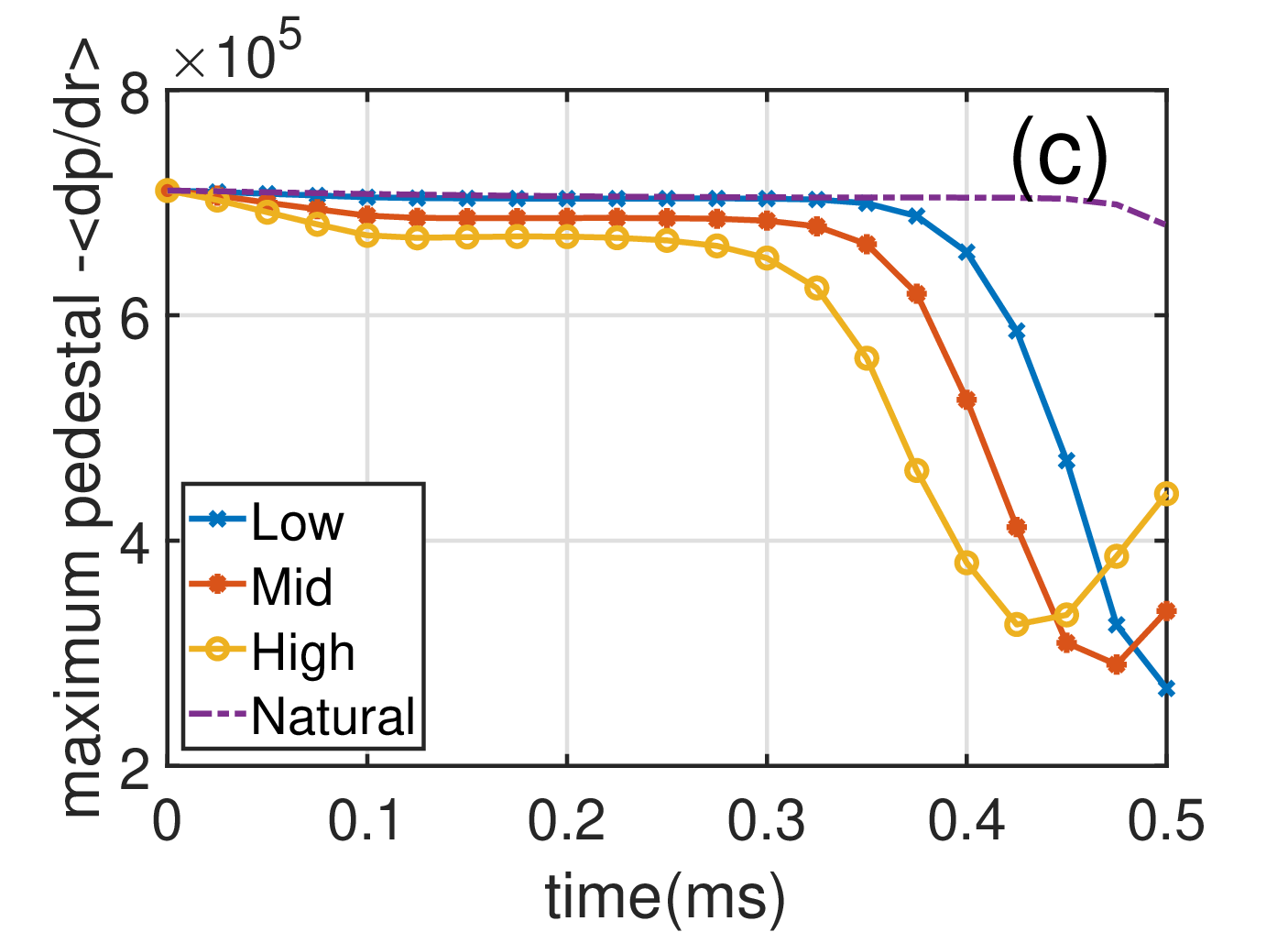}
	\end{center}
	\caption{(a) Pedestal pressure profiles on the outboard mid-plane at $t=0.1$ ms are shown for various impurity density levels. (b) Local pressure gradient at the pedestal foot region on the outboard mid-plane and (c) the maximum flux-surface-averaged pedestal pressure gradient are plotted as functions of time for various impurity density levels.}
	\label{fig: impurity resistivity-level high n mode and pres}
\end{figure}

\begin{figure}[ht]
	\begin{center}
		\includegraphics[width=0.55\linewidth]{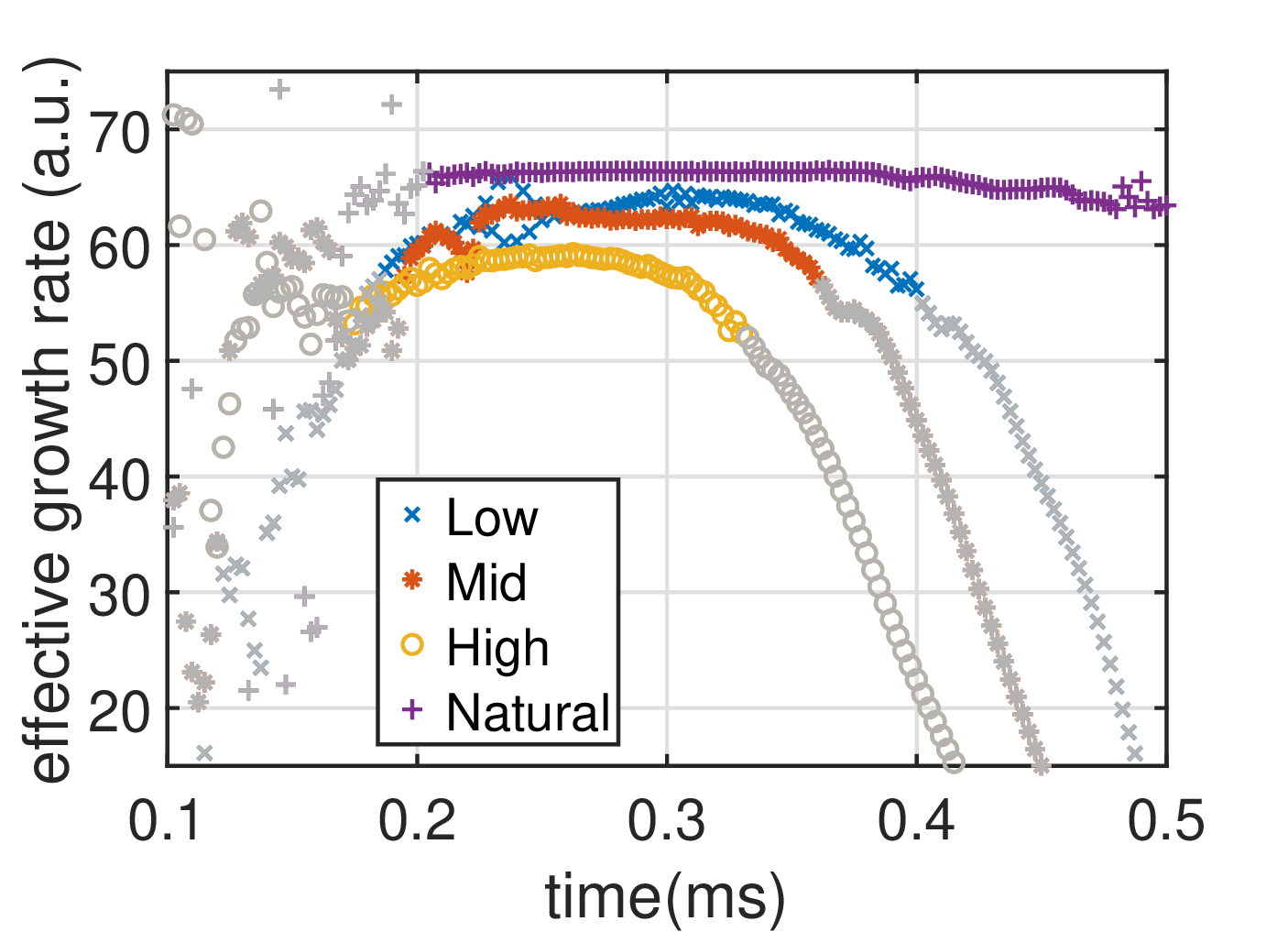}
	\end{center}
	\caption{Effective growth rates of the $n=21$ mode ($\rm{d}\ln \textit{W}_{mag,n=21}/\rm{d}\textit{t}$, where $W_{mag,n=21}$ denotes the magnetic energy of the $n=21$ mode) are plotted as a function of time for various impurity density levels.}
	\label{fig: impurity resistivity-level growth rate dpdr ave}
\end{figure}

Impurity seeding at the plasma separatrix modifies the pedestal plasma pressure profile, inducing a bump at the pedestal foot that amplifies with higher impurity density levels (Fig. \ref{fig: impurity resistivity-level high n mode and pres}a). On the one hand, this perturbation shifts the pressure profile outward and enhances the local pressure gradient at the outboard mid-plane pedestal foot, and such effects intensify at higher impurity density levels (Fig. \ref{fig: impurity resistivity-level high n mode and pres}b). On the other hand, the enhanced pressure at the pedestal foot mitigates the steep gradient at the pedestal waist, thereby reducing the maximum flux-surface-averaged pedestal pressure gradient at that pedestal location. Higher impurity density levels amplify this reduction (Fig. \ref{fig: impurity resistivity-level high n mode and pres}c). Consequently, these impurity seeding induced pedestal pressure perturbations primarily govern the $n=21$ mode growth and ELM dynamics.
The observed correlation between the higher impurity density and the faster ELM crash onset originates from the accelerated early-stage mode growth ($t<0.1$ ms in Fig. \ref{fig: impurity resistivity-level lem size}b1). Earlier mode saturation enables earlier ELM triggering, as exponential growth rates remain comparable across cases. The rapid initial growth ($t<0.1$ ms) is driven by the enhanced local pressure gradient during impurity density ramping, with higher impurity density levels leading to earlier growth (Fig. \ref{fig: impurity resistivity-level high n mode and pres}b). In contrast, in early stage the mode decays in the reference case, where the impurity induced perturbations are absent.
The $n=21$ mode's exponential growth rate is proportional to the maximum flux-surface-averaged pressure gradient in the pedestal region, a parameter that decreases with increasing impurity density (Fig. \ref{fig: impurity resistivity-level high n mode and pres}c and Fig. \ref{fig: impurity resistivity-level growth rate dpdr ave}). Therefore, higher impurity density levels suppress the exponential growth rate, yielding smaller mode amplitude peaks and ELM sizes.

Numerous experiments demonstrate that the amount of injected material plays a critical role in ELM dynamics, where higher impurity densities increase ELM frequency with reduced ELM size. For example, this trend is observed in pellet ELM pacing experiments on D\uppercase\expandafter{\romannumeral3}-D \cite{Baylor_PhysRevLett.110.245001,Baylor_2013}, JET \cite{Lang_2013} and ASDEX-Upgrade \cite{Lang_2004}, and SMBI impurity seeding experiments on HL-2A \cite{Xiao_2012}. The simulation results are consistent with these experimental findings.




\subsection{The poloidal location of impurity seeding}
\label{subsection: Impurity injection geometry}

\begin{figure}[ht]
	\begin{center}
		\includegraphics[width=0.45\linewidth]{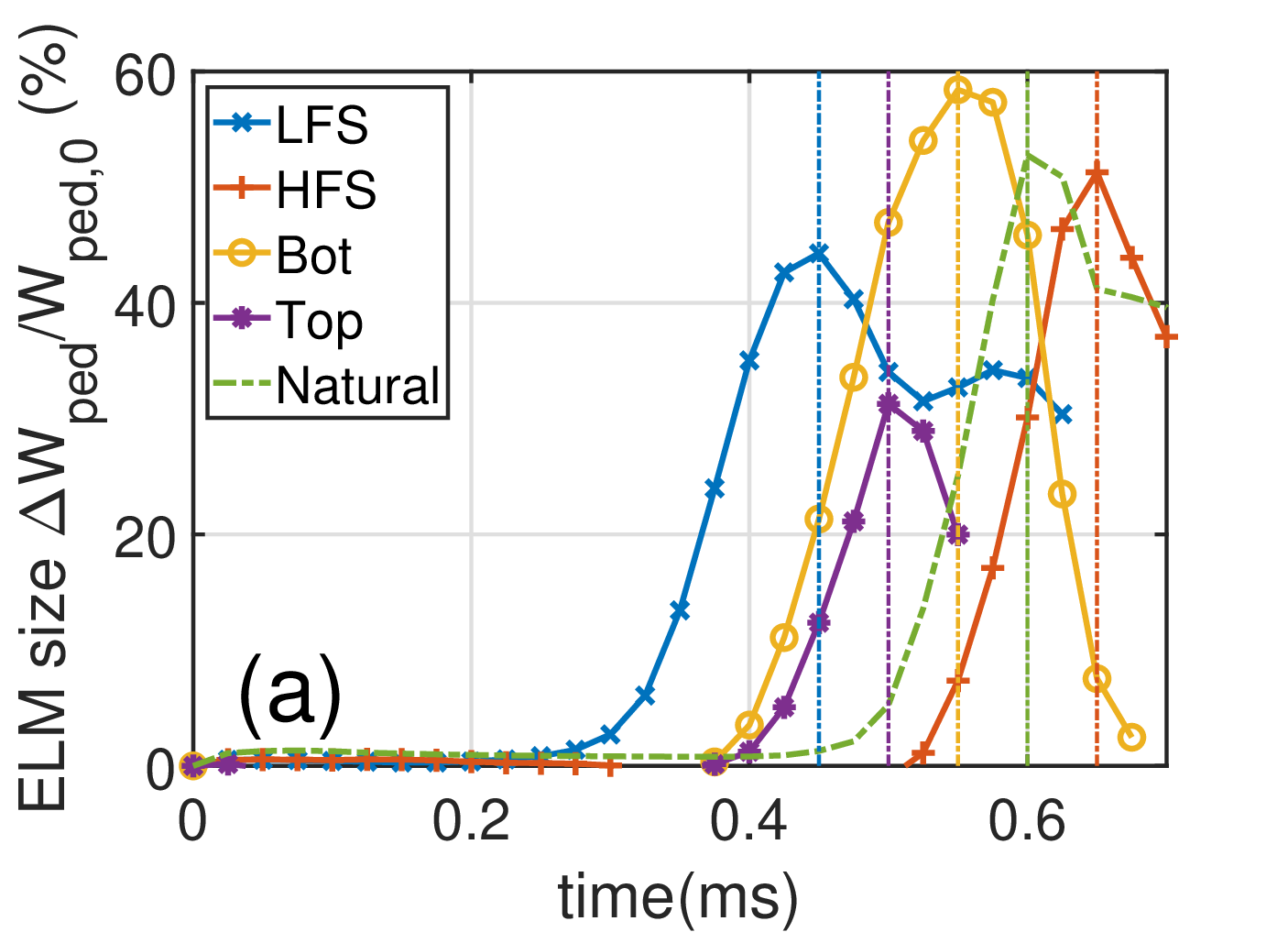}
		\includegraphics[width=0.45\linewidth]{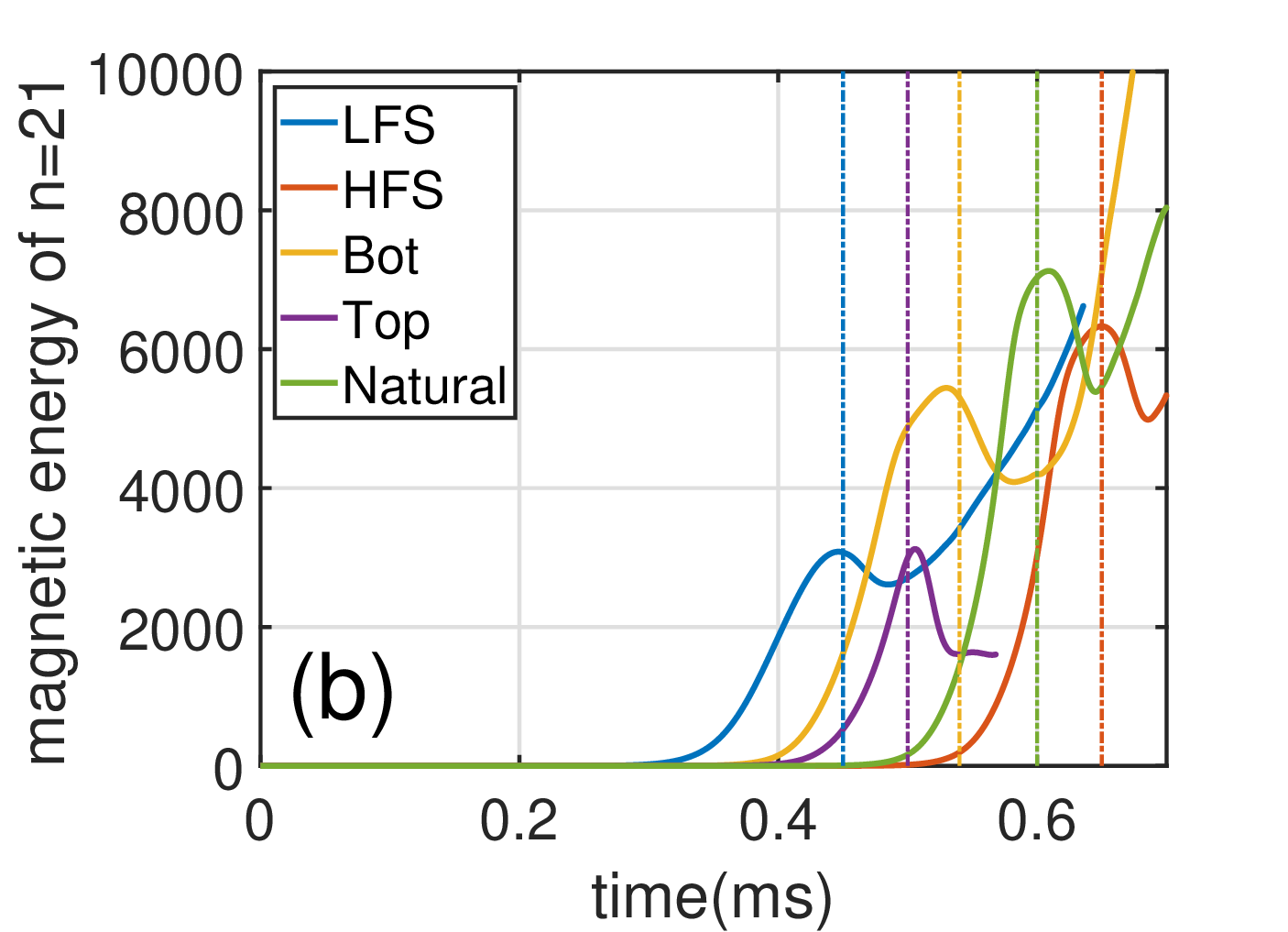}
	\end{center}
	\caption{(a) ELM size and (b) magnetic energy of the $n=21$ mode are shown as functions of time for different poloidal impurity seeding locations. Labels `LFS', `HFS', `Bot', and `Top' correspond to impurity seeding at the outboard mid-plane, inboard mid-plane, lower X-point, and top of the device, respectively. `Natural' denotes the reference case without impurity seeding. The Lundquist number in the plasma core is set to $S=1\times10^9$.}
	\label{fig: impurity resistivity-location lem size}
\end{figure}

Fig. \ref{fig: impurity resistivity-location lem size}(a) compares the ELM crashes with various poloidal impurity seeding locations, and the initial impurity density level remains identical across all cases. Both the ELM crash onset time and the resulting ELM size $\Delta W_{ped}/\Delta W_{ped,0}$ depend strongly on the poloidal seeding location. The `LFS' impurity seeding triggers ELM crash most effectively, resulting in the fastest crash, whereas the `HFS' seeding delays the crash the most. Moreover, the `Top' and the `Bot' seeding trigger ELM crash at similar time. The `Top' seeding yields the smallest ELM size, whereas the `Bot' seeding produces the largest. Despite variations in nonlinear ELM dynamics, the $n=21$ mode remains the most unstable mode and shows a strong correlation with the ELM crash evolution (Fig. \ref{fig: impurity resistivity-location lem size}b). The ELM size peak aligns well with the mode's amplitude peak in time across all cases. Furthermore, the maximum ELM size generally scales with the amplitude of this peak, except for the `Bot' and the `Top' seeding cases.

\begin{figure}[ht]
	\begin{center}
		\includegraphics[width=0.5\linewidth]{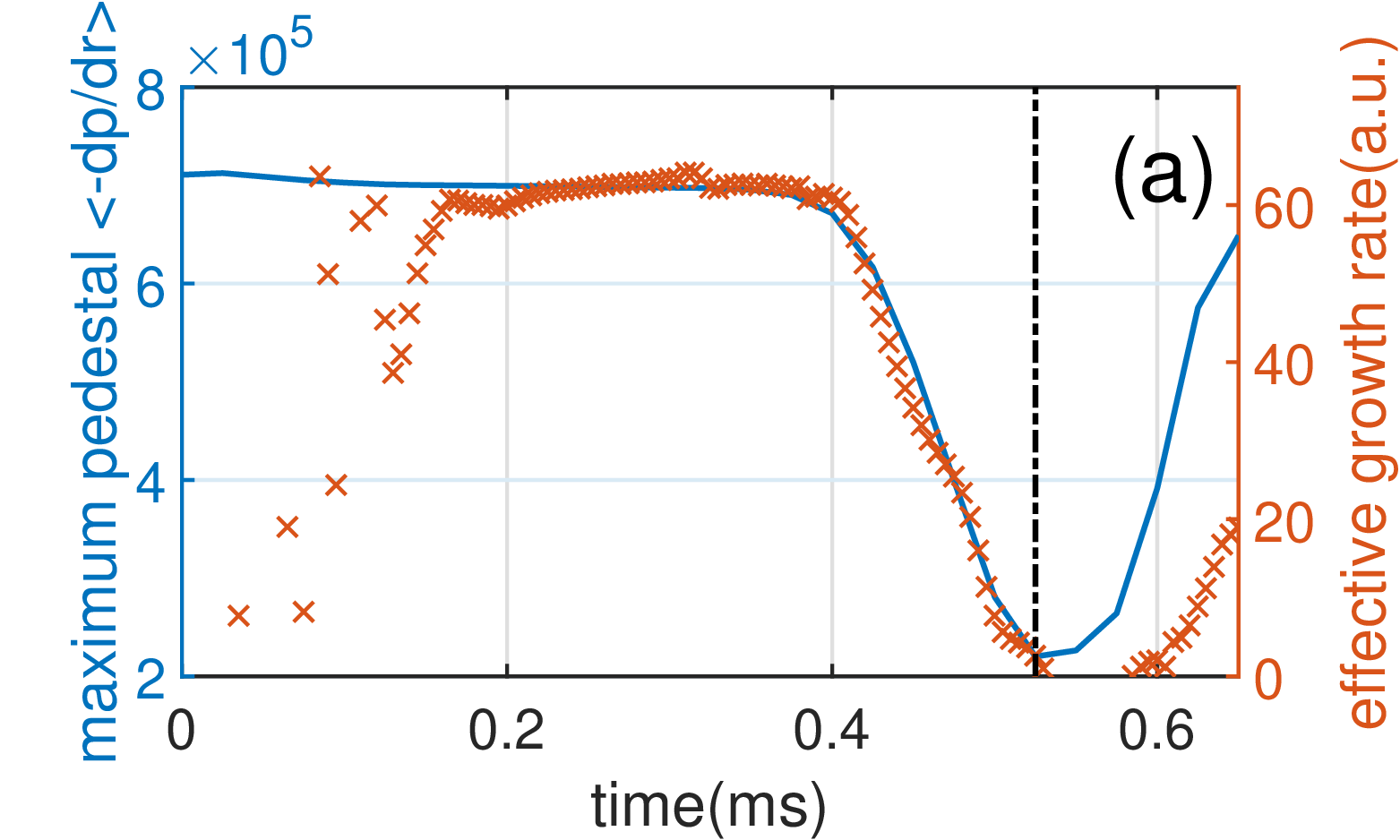}
		\includegraphics[width=0.35\linewidth]{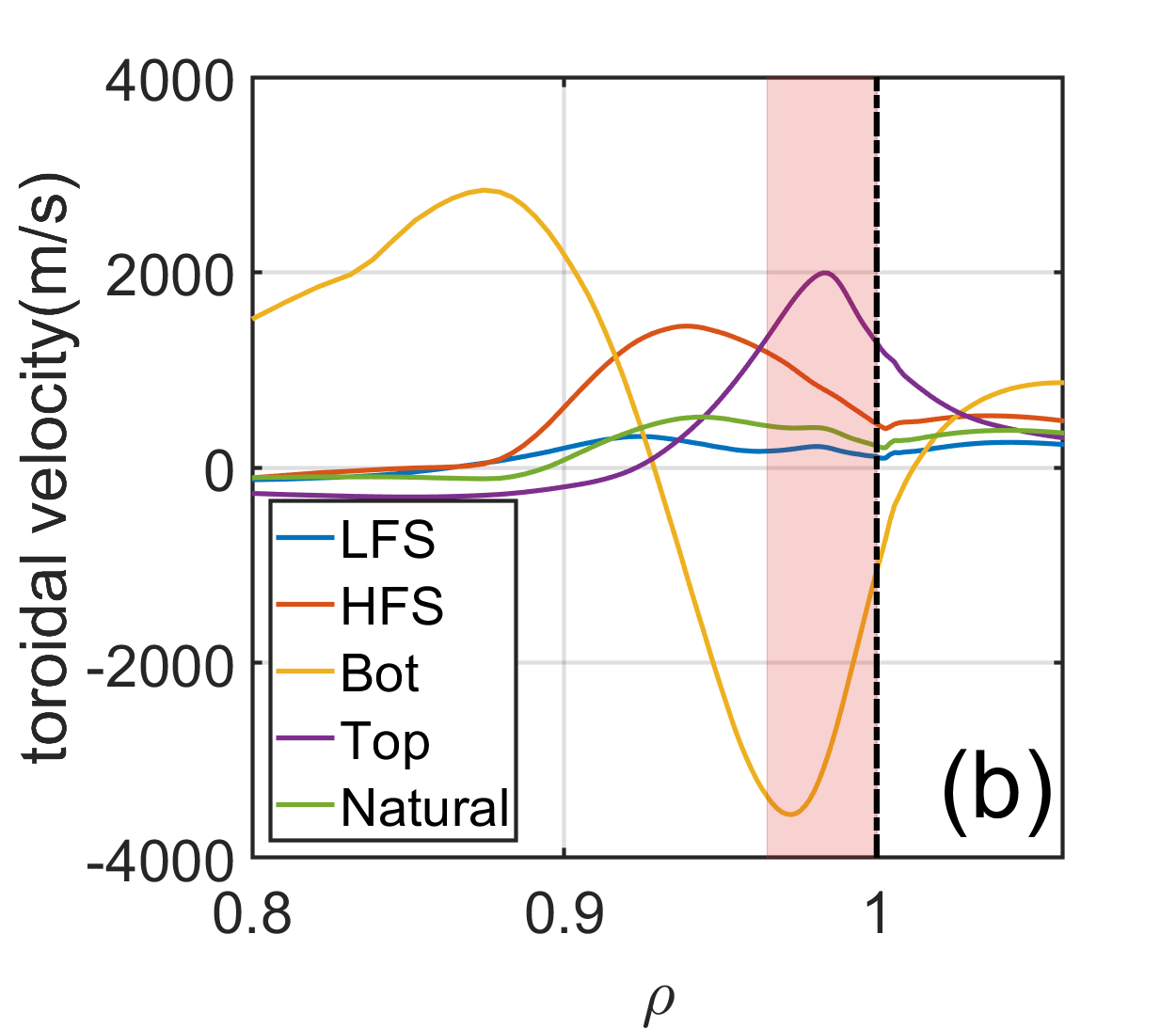}
	\end{center}
	\caption{(a) Effective growth rate of the $n=21$ mode ($\rm{d}\ln \textit{W}_{mag,n=21}/\rm{d}\textit{t}$) and the maximum flux-surface-averaged pedestal pressure gradient are plotted as functions of time for the 'Bot' impurity seeding case. (b) Radial profiles of flux-surface-averaged toroidal velocity for various poloidal impurity seeding locations at the time of maximum ELM size.}
	\label{fig: impurity resistivity-bot growth}
\end{figure}

\begin{figure}[ht]
	\begin{center}
		\includegraphics[width=0.45\linewidth]{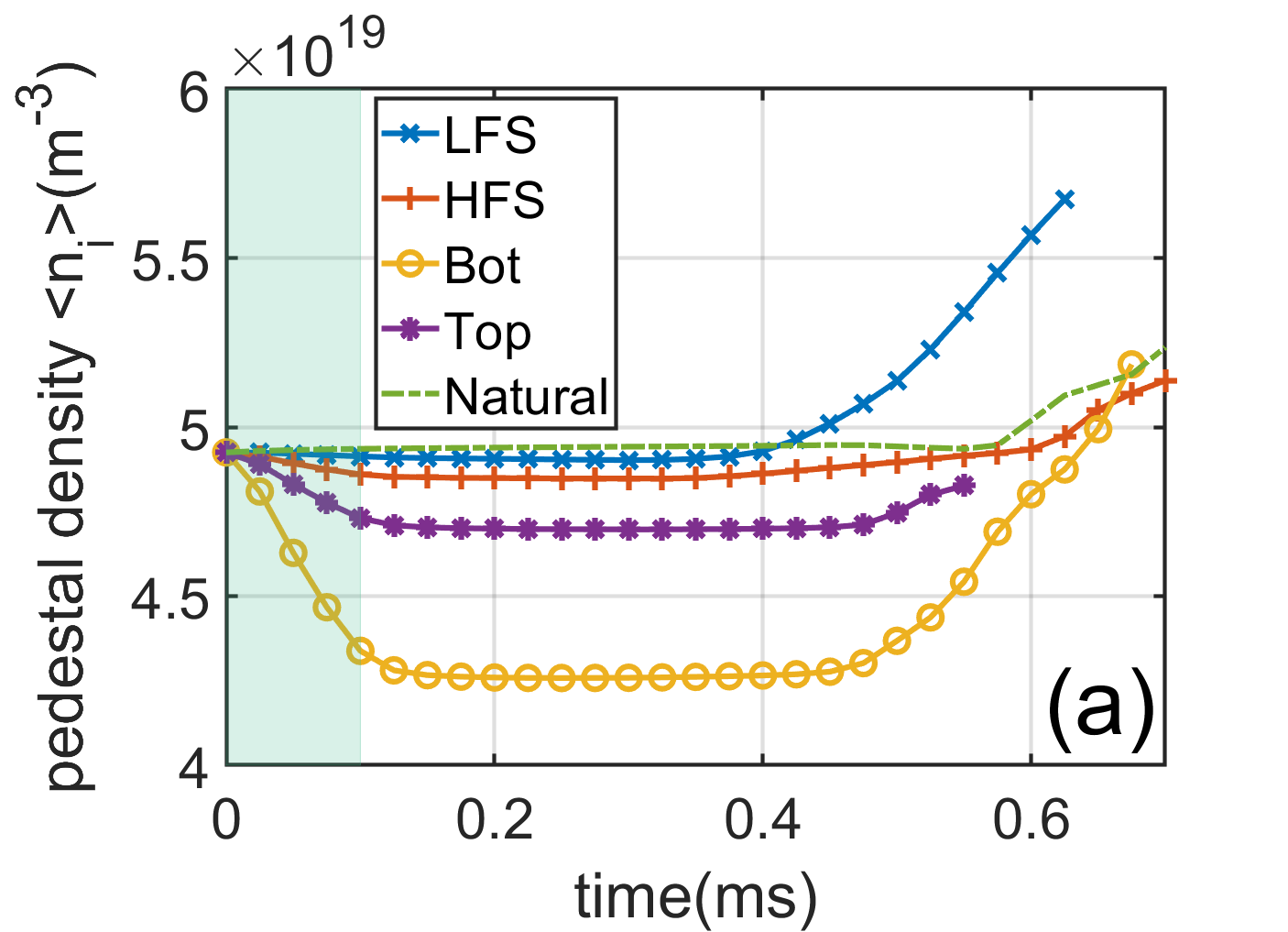}
		\includegraphics[width=0.45\linewidth]{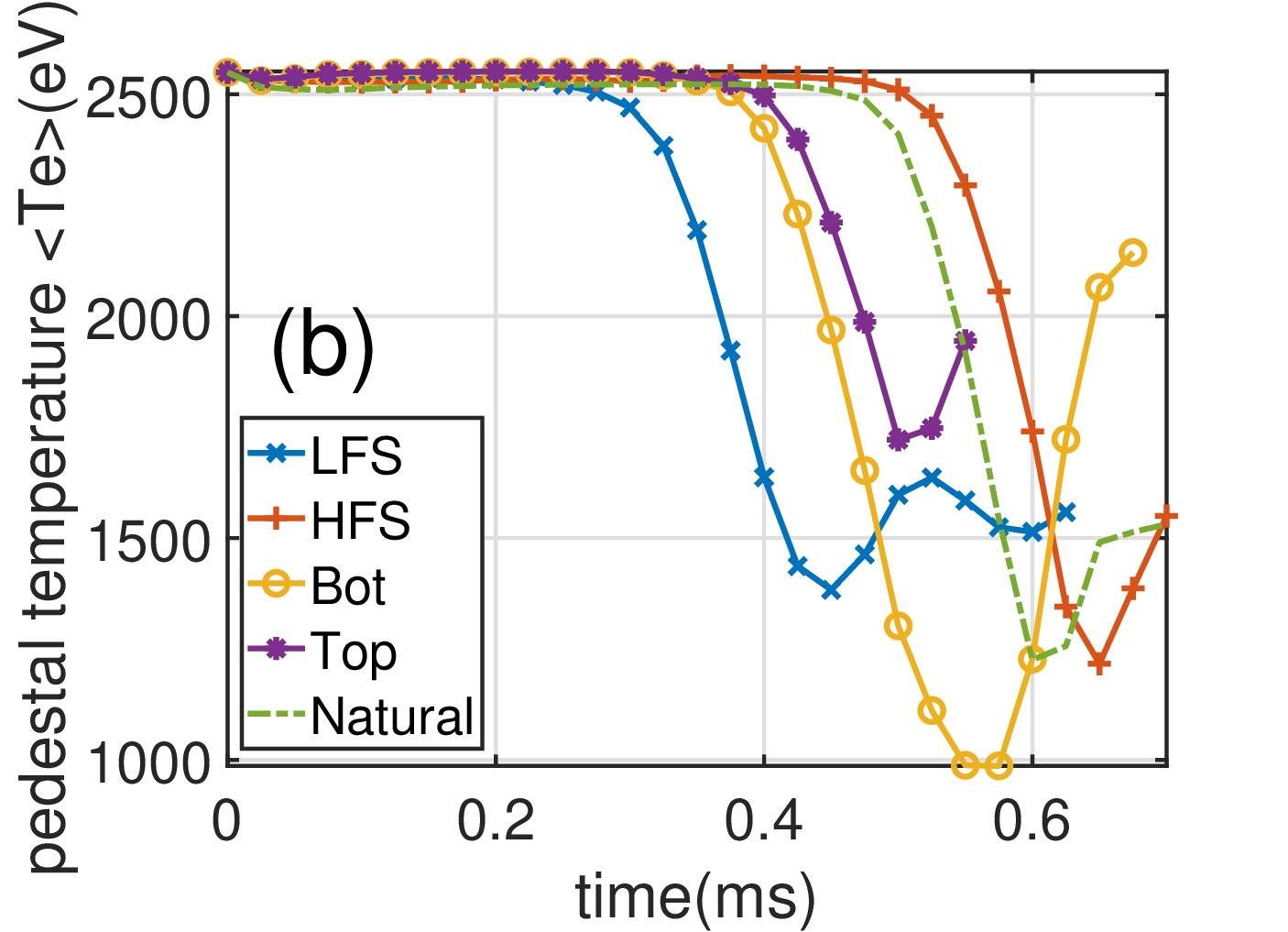}
	\end{center}
	\caption{Time evolution of the pedestal-averaged (a) plasma ion density $\left\langle n_i\right\rangle =\int_{S_{ped}}n_i\rm d\textit{S}/\textit{S}_{ped}$ and (b) electron temperature $\left\langle T_e\right\rangle =\int_{S_{ped}}T_e\rm d\textit{S}/\textit{S}_{ped}$ for various poloidal impurity seeding locations, where $S_{ped}$ denotes the poloidal cross-section area of the plasma pedestal region.}
	\label{fig: impurity resistivity-bot top Te ni loss}
\end{figure}

\begin{figure}[ht]
	\begin{center}
		\includegraphics[width=0.45\linewidth]{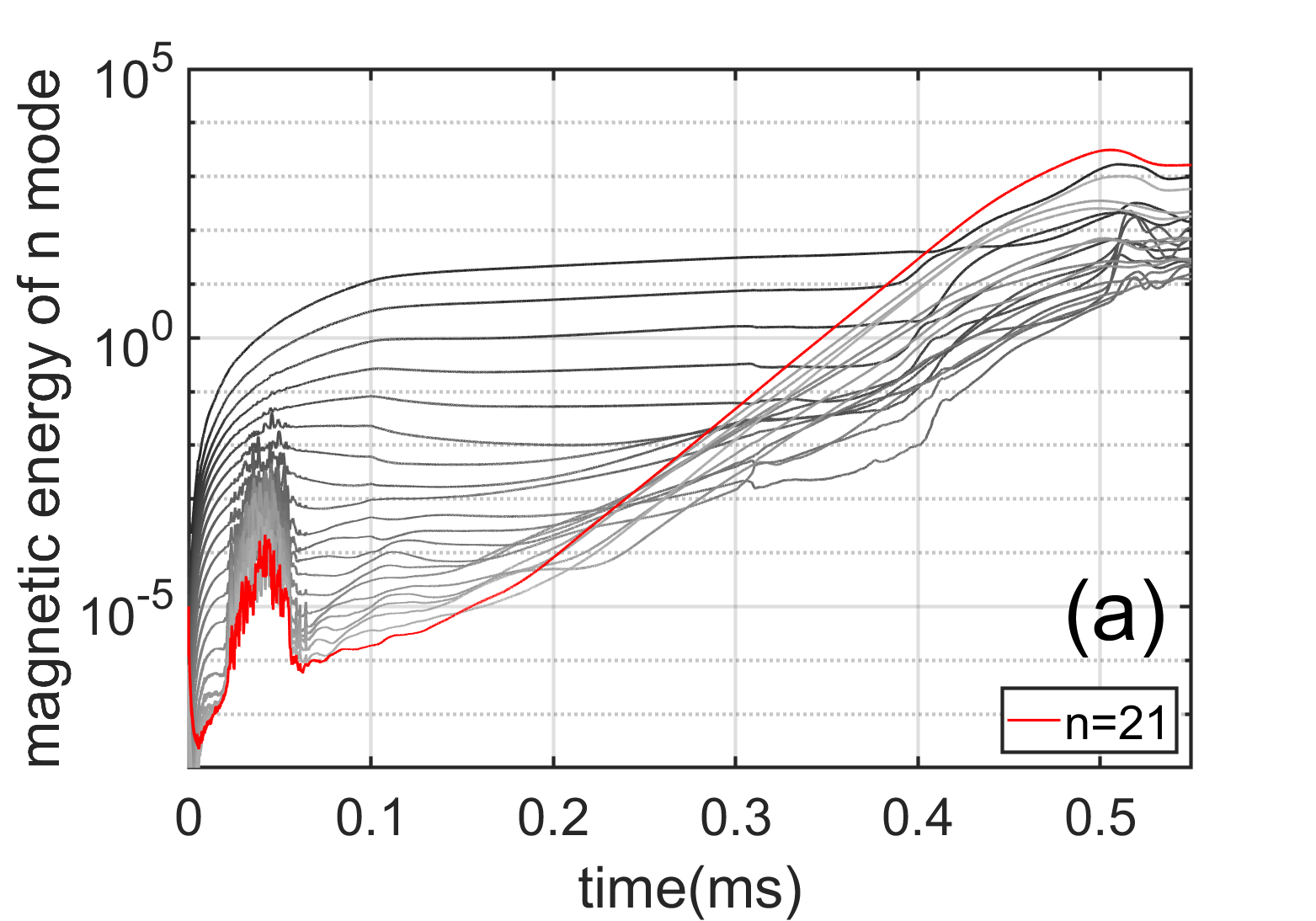}
		\includegraphics[width=0.45\linewidth]{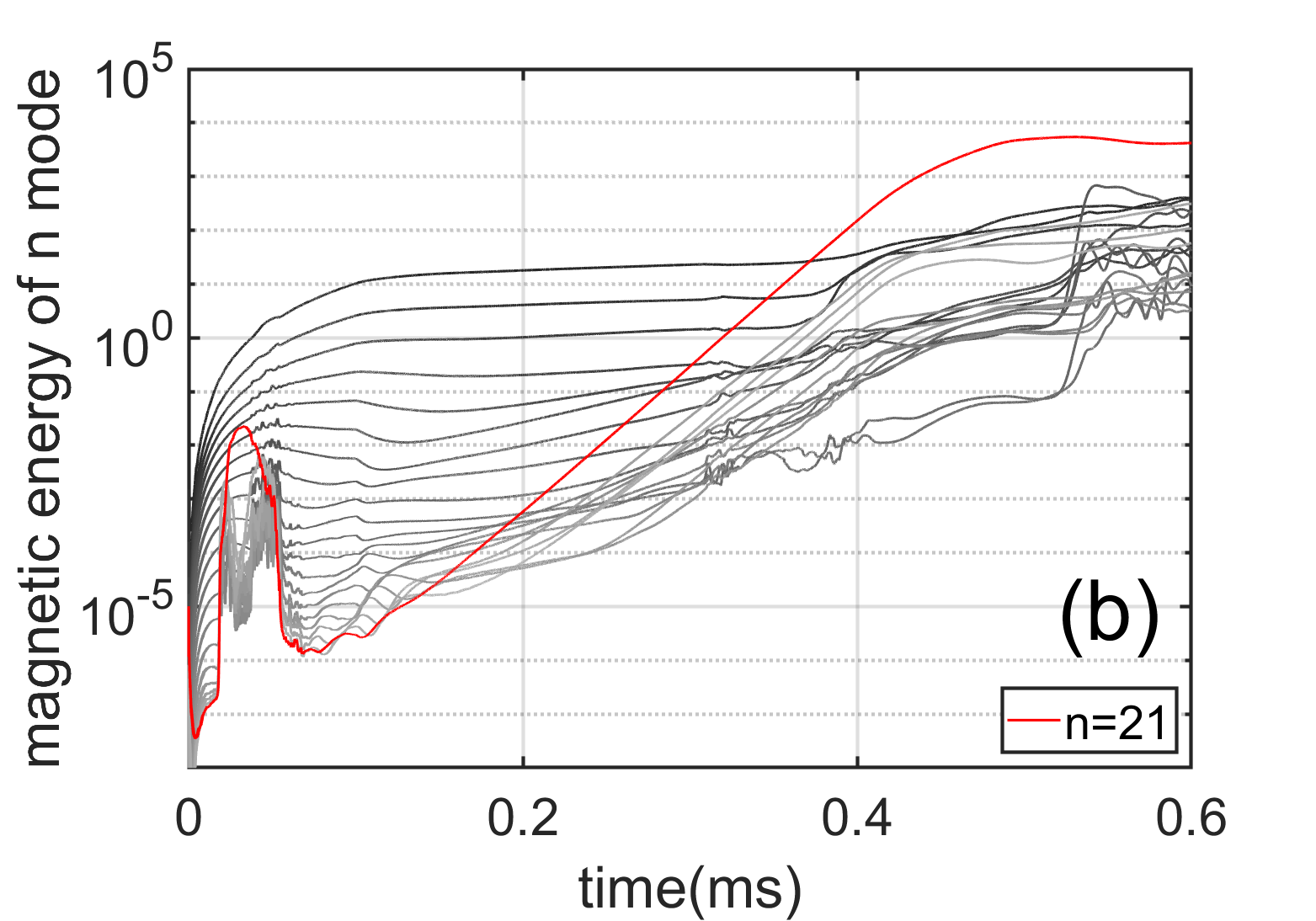}
	\end{center}
	\caption{Magnetic energy evolution of selected toroidal modes are shown as functions of time for (a) the `Top' and (b) the 'Bot' impurity seeding cases, with the $n=21$ mode (red line) highlighted.}
	\label{fig: impurity resistivity-bot top max mangetic }
\end{figure}

\begin{figure}[ht]
	\begin{center}
		\includegraphics[width=0.45\linewidth]{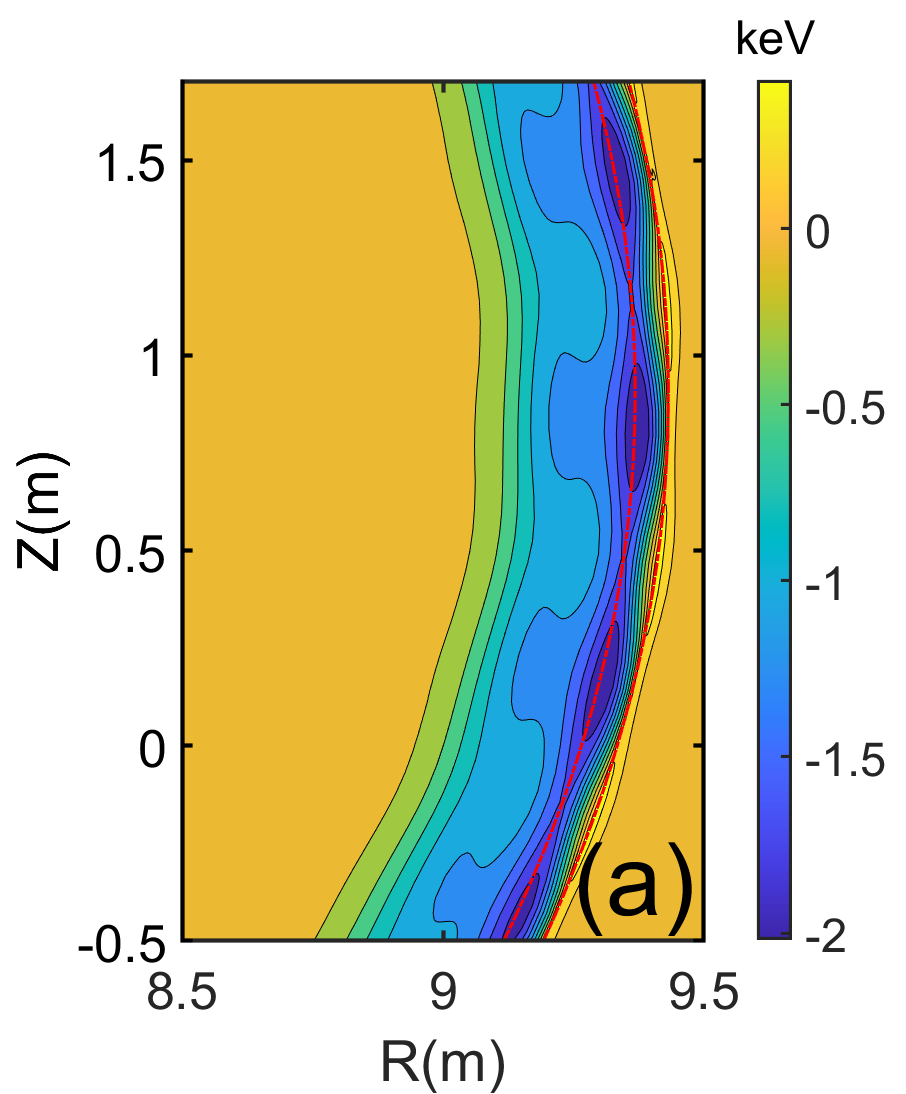}
		\includegraphics[width=0.45\linewidth]{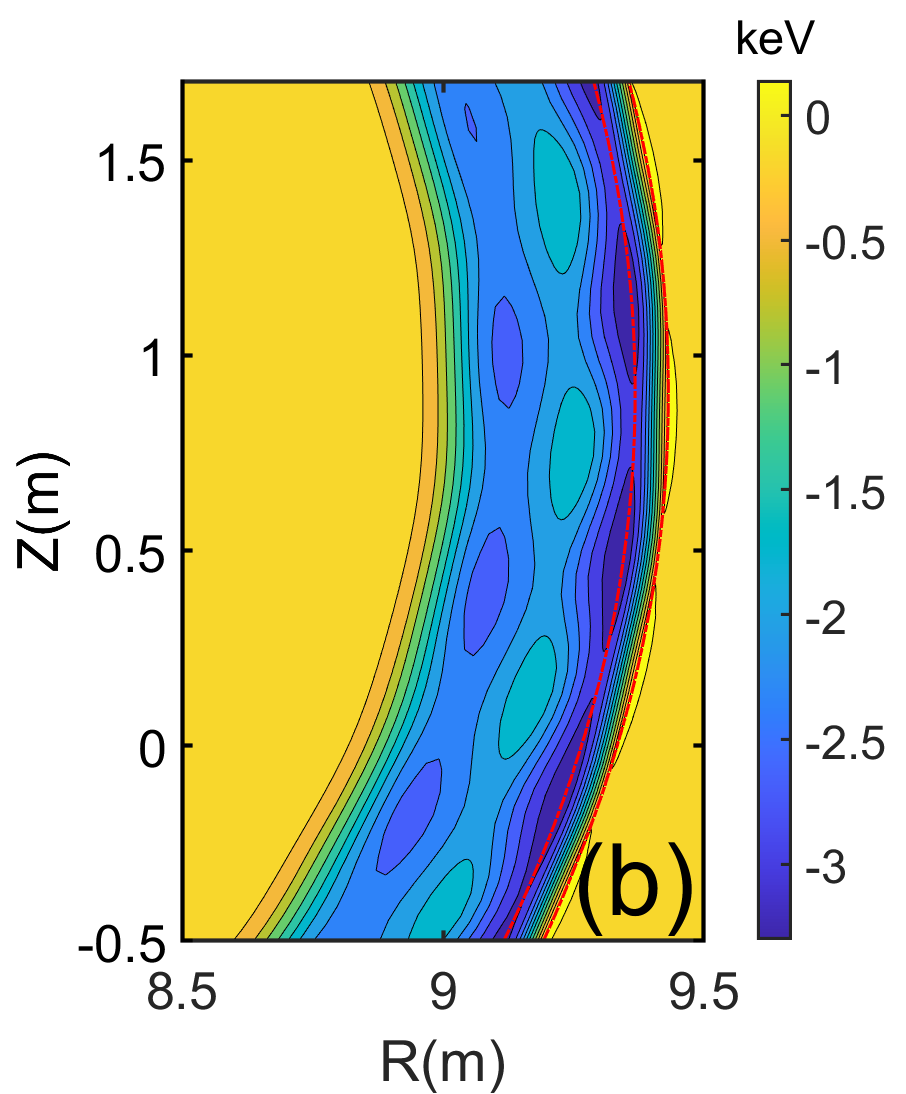}
	\end{center}
	\caption{Electron temperature perturbation contours for (a) the `Top' seeding at $t=0.5$ ms and (b) the 'Bot' seeding at $t=0.55$ ms, with the pedestal top and plasma separatrix marked using red dashed lines.}
	\label{fig: impurity resistivity-bot top filaments}
\end{figure}

The `Bot' impurity seeding leads to the largest ELM size, indicating that the corresponding saturation peak of the $n=21$ mode should be the highest. Whereas the $n=21$ mode growth is primarily driven by the pedestal pressure gradient (e.g., the `Bot' seeding case; Fig. \ref{fig: impurity resistivity-bot growth}a), the `Bot' seeding case develops a strong flow that exceeds those from other seeding locations at the time of maximum ELM size (Fig. \ref{fig: impurity resistivity-bot growth}b). This intense flow and its associated shear likely suppress the mode growth rate and reduce the peak amplitude. Furthermore, sheared toroidal rotation is known to stabilize PBMs \cite{Saarelma_2007}, and similar flow dynamics has been found in JOREK simulations \cite{Huysmans_2007}. The largest ELM size in the `Bot' case is linked to its impurity injection location. Compared to other locations, the `Bot' seeding at the lower X-point in this single-null plasma configuration induces stronger separatrix perturbations, thereby enhancing confinement degradation. This is corroborated by the largest density loss in the `Bot' case during the initial $0$--$0.1$ ms phase of impurity density ramping (Fig. \ref{fig: impurity resistivity-bot top Te ni loss}a),
as well as the pedestal temperature drop during the ELM crash (Fig. \ref{fig: impurity resistivity-bot top Te ni loss}b). Additionally, the pedestal density increase during the crash stems mainly from the outward core plasma fluxes (Fig. \ref{fig: impurity resistivity-bot top Te ni loss}a).
The smallest ELM size observed in the `Top' seeding case can be explained with the mode growth. While the $n=21$ mode remains dominant in all cases (e.g., the `Bot' seeding case; Fig. \ref{fig: impurity resistivity-bot top max mangetic }b), its dominance diminishes in the `Top' seeding case (Fig. \ref{fig: impurity resistivity-bot top max mangetic }a). A single dominant high-$n$ mode growth causes more intense perturbations across a broader plasma edge region (Fig. \ref{fig: impurity resistivity-bot top filaments}b). These perturbations lead to the development of adjacent and significantly overlapping multiple island-chain structures in the edge temperature, which are likely the outcome of magnetic field line stochasticity, thereby facilitating ELM energy loss. In contrast, high-$n$ mode growth is suppressed in the `Top' case, with subsequent perturbations limited in scope (Fig. \ref{fig: impurity resistivity-bot top filaments}a). This suppression probably stems from mode competition among growth-comparable modes, resulting in reduced ELM size.
Besides, both the `Top' and `Bot' cases exhibit an initial transient peak affecting high-$n$ modes ($n\ge12$), which originates from early-stage density perturbations (Figs. \ref{fig: impurity resistivity-bot top max mangetic }).

\begin{figure}[ht]
	\begin{center}
		\includegraphics[width=0.55\linewidth]{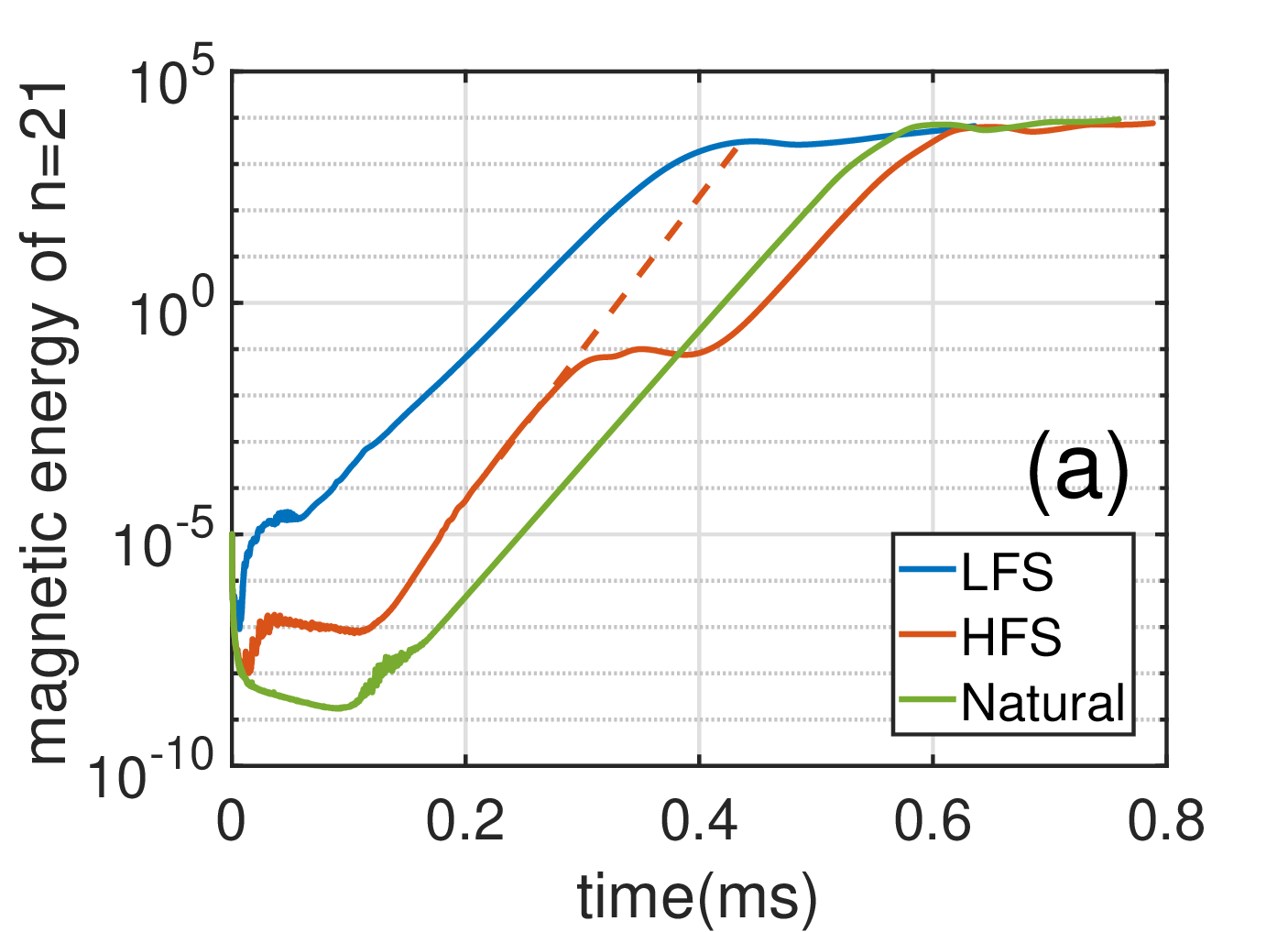}
		\includegraphics[width=0.45\linewidth]{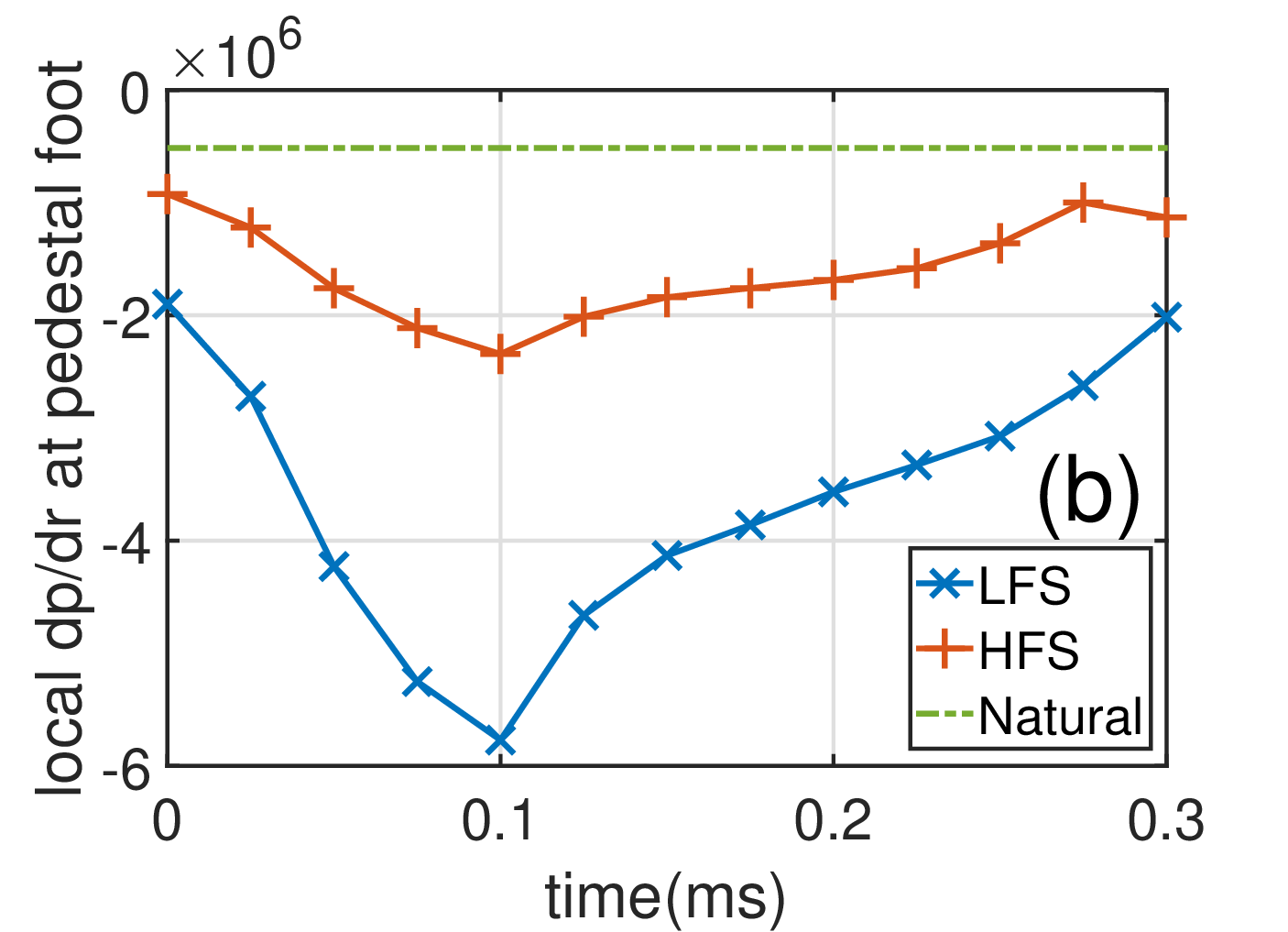}
		\includegraphics[width=0.38\linewidth]{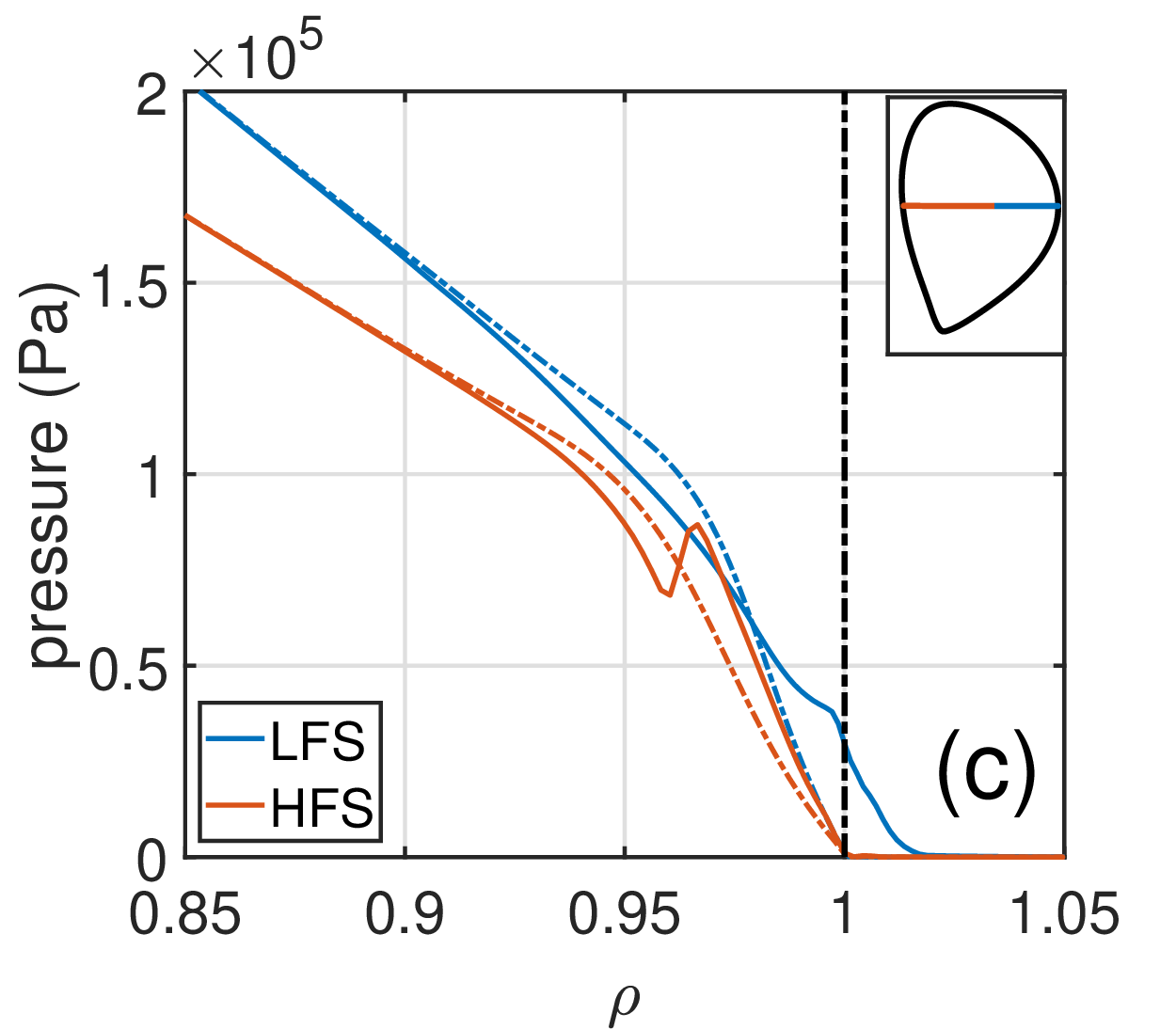}
	\end{center}
	\caption{(a) Magnetic energy of the $n=21$ mode and (b) local pressure gradient at the pedestal foot region are plotted as functions of time for different poloidal impurity seeding locations. (c) Radial pressure profiles at $t=0.3$ ms along the correspondingly colored radial lines in the top-right inset, with the initial equilibrium represented by dashed lines and the plasma separatrix marked using a black dashed line.}
	\label{fig: impurity resistivity-lfs hfs magnetic n=21 and pres}
\end{figure}

The growth of the $n=21$ mode, primarily driven by impurity modified pedestal pressure profile, plays the central role in ELM crash onset for the `LFS' and the `HFS' impurity seeding cases (Fig. \ref{fig: impurity resistivity-lfs hfs magnetic n=21 and pres}a). In both cases, impurity seeding enhances the local pedestal pressure gradient during the initial $0.1$ ms of impurity density ramping (Fig. \ref{fig: impurity resistivity-lfs hfs magnetic n=21 and pres}b). The `LFS' seeding localizes pressure perturbations on the bad-curvature side, directly destabilizing ballooning mode instabilities. Conversely, the `HFS' seeding induces perturbations on the good-curvature side with a delayed early-stage growth (Fig. \ref{fig: impurity resistivity-lfs hfs magnetic n=21 and pres}a). Notably, if the mode's exponential growth rate persists until saturation (dashed line in Fig. \ref{fig: impurity resistivity-lfs hfs magnetic n=21 and pres}a), the `HFS' seeding would trigger earlier ELM crash than the reference case. However, a secondary stagnation phase develops in the `HFS' case between $t=0.3$ and $0.4$ ms, coinciding with a localized pressure profile peak developed in the pedestal region (Fig. \ref{fig: impurity resistivity-lfs hfs magnetic n=21 and pres}c). The mode resumes exponential growth after the mitigation of this peak, and the growth continues until saturation. This pressure profile peak reduces the maximum flux-surface-averaged pedestal pressure gradient, thereby suppressing the exponential mode growth. The `HFS' seeding induced pressure peak arises from deeper impurity penetration and subsequent ionization, probably due to the inward shift of the local dense perturbations on the high-field side. Similar deep penetration effects are also observed in ASDEX Upgrade experiments and are attributed to toroidal curvature effects \cite{Lang1997PRL}.

\section{Discussion}
\label{Sec: discussion}

In D\uppercase\expandafter{\romannumeral3}-D experiments, outboard mid-plane pellet triggered ELMs exhibit lower energy loss compared to those triggered by the X-point pellet injection \cite{Baylor_2013}, in agreement with the `LFS' and the `Bot' impurity seeding simulation results reported here. JET experiments demonstrate that the vertical high-field side (near the device top) pellet injection achieves higher ELM triggering probability than outboard mid-plane injection \cite{Lang_2013,Lennholm_2021}. Our simulations find that the `Top' seeding delays ELM crash relative to the `LFS' seeding. However, higher ELM triggering probability does not necessarily correlate with faster ELM crash.
Systematically exploring ELM triggering efficiency across a broad range of impurity injection locations remains experimentally challenging due to spatial constraints on impurity seeding. JOREK simulations indicate enhanced ELM triggering from the high-field side compared to the low-field side when considering ballooning modes only without peeling-ballooning coupling \cite{Futatani_2014}. In contrast, nonlinear M3D-C1 simulations identify the outboard mid-plane as the most effective location for ELM triggering, with lower/upper X-point regions being least favorable \cite{Wingen_2024}. Compared to those studies, our findings suggest the outboard mid-plane as the optimal impurity seeding location for ELM triggering. Ultimately, impurity seeding location emerges as a critical parameter for ELM mitigation, though definitive conclusions remain elusive due to variations in impurity seeding models and plasma equilibrium scenarios across studies, warranting further investigation.

JOREK multi-ELM cycle simulations reveal that the first ELM crash, initiated by arbitrary seed perturbations, exhibits larger energy loss than subsequent crashes driven by self-consistent perturbations \cite{Orain_PhysRevLett.114.035001,Cathey_2020}. This occurs because the initial equilibrium is more susceptible to PBM instabilities than later pre-ELM profiles, as the plasma struggles to fully recover its initial state in simulations. Consequently, the energy released during the first ELM crash provides an upper limit for evaluating ELM mitigation strategies. The single-ELM simulations studied here represent this most hazardous scenario.
Furthermore, determining the maximum achievable impurity seeded ELM frequency is crucial, since energy per ELM decreases with higher frequency. For example, in the CFETR hybrid scenario ($\sim40$ MJ energy release), the impurity seeded ELM frequency must increase by at least $40$ times compared to the natural ELM frequency. Our simulations have demonstrated that higher impurity density levels accelerate ELM crash while reducing its size. The ELM frequency upper limit may be determined by the mode growth rate, which triggers ELM crash as MHD instabilities approach saturation and is strongly modulated by impurity induced pedestal profile modifications. Developing multi-ELM cycle simulation is essential for reliable ELM mitigation predictions, a key focus of future work.


\section{Conclusion}
\label{Sec: conclusion}

In this work, we study the nonlinear dynamics of ELM crashes under various impurity seeding scenarios using NIMROD simulations. The characteristic features of ELM crash---filamentary structures, plasma pedestal collapse, and energy loss---are reproduced and agree with experimental observations. The high-$n$ ballooning mode instability, primarily driven by the pedestal pressure gradient, is identified as the most unstable mode and the dominant trigger of ELM crash in this pressure-limited pedestal equilibrium.
Impurity seeding significantly influences ELM dynamics by modifying the pedestal pressure profile. The perturbed local pressure gradient drives early-stage mode growth, whereas the subsequent exponential growth of the high-$n$ ballooning mode directly correlates with the maximum flux-surface-averaged pressure gradient. Consequently, higher impurity density levels accelerate ELM crash onset and reduce ELM size. Furthermore, the poloidal location of impurity seeding plays a critical role in ELM crash onset and size: the outboard mid-plane seeding is identified as the most effective trigger for ELM crash, and the lower X-point seeding produces the largest ELM size.

\section*{Acknowledgments}
We are grateful for the supports from the NIMROD team and the CFETR physics design team. This work was supported by the National Magnetic Confinement Fusion Program of China (Grant No. 2019YFE03050004), Hubei International Science and Technology Cooperation Projects (No. 2022EHB003), and U.S. Department of Energy (Grant No. DE-FG02-86ER53218).
The computing work in this paper was supported by the Public Service Platform of High Performance Computing by Network and Computing Center of HUST, and this research used resources of the National Energy Research Scientific Computing Center, a DOE Office of Science User Facility supported by the Office of Science of the U.S. Department of Energy under Contract No. DE-AC02-05CH11231 using NERSC award FES-ERCAP0027638.

\newpage
\section*{References}

\bibliographystyle{iopart-num}
\bibliography{elm}

\newpage



\newpage

\newpage

\newpage

\newpage

\newpage

\newpage

\newpage

%

\newpage

\newpage

\newpage

\newpage

\end{document}